\documentclass[a4paper, 11pt]{article}
\usepackage[hmargin=2.5cm,vmargin=3.5cm]{geometry}
\usepackage[utf8]{inputenc}
\usepackage{amsmath}
\usepackage{booktabs} 
\usepackage{caption}   
\usepackage{subcaption}  
\usepackage{mathrsfs}

\usepackage{graphicx}
\usepackage{hyperref}
\usepackage{multicol}
\usepackage{mathtools} 
\usepackage{eurosym}
\usepackage{pdfpages}
\usepackage[normalem]{ulem}
\usepackage{latexsym}
\usepackage{amssymb}
\usepackage[capposition=top]{floatrow}
\usepackage{lipsum}
\usepackage{setspace}
\usepackage{float}
\usepackage[english]{babel}
\usepackage{fancyhdr, afterpage}
\usepackage{bbm}
\usepackage{amsfonts}
\usepackage{bm}
\usepackage{pdflscape} 
\usepackage{tikz}
\usepackage{csquotes}
\usepackage[subnum]{cases}
\DeclareMathOperator*{\E}{\mathbb{E}}
\hypersetup{
  colorlinks=true,
  linkcolor=blue,
  filecolor=black, 
  urlcolor=blue,
  citecolor=blue
}
\usepackage[authordate,bibencoding=auto,backend=biber,natbib,maxcitenames=1,uniquename=false,
  uniquelist=false,doi=false,isbn=false,url=false,eprint=false]{biblatex-chicago}
\renewcommand{\footnoterule}{%
  \kern -3pt
  \hrule width \textwidth height 1pt
  \kern 2pt
}
\title{Firm Heterogeneity and Macroeconomic Fluctuations: a Functional VAR model\thanks{Andrea Renzetti and Massimiliano Marcellino thank MUR-Prin 2022 - Prot. 20227YZ9JK, financed by the European Union - Next Generation EU, for partial financial support. Seminar participants at Bocconi University provided useful comments on a previous draft. We thank Joshua C. Chan for helpful comments.}}
\author{Massimiliano Marcellino \thanks{ Bocconi University and BAFFI-CAREFIN Center; email:  \href{mailto:massimiliano.marcellino@unibocconi.it}{massimiliano.marcellino@unibocconi.it}} \and Andrea Renzetti \thanks{Bocconi University and BAFFI-CAREFIN Center; email: \href{mailto:andrea.renzetti@unibocconi.it}{andrea.renzetti@unibocconi.it}.} \and Tommaso Tornese \thanks{Universitá Cattolica del Sacro Cuore, Milano; email: \href{mailto:tommaso.tornese@unicatt.it}{tommaso.tornese@unicatt.it}}}
\date{ This Draft: \today \\
}

\begin{document}
\maketitle
\begin{abstract}
We develop a Functional Augmented Vector Autoregression (FunVAR) model to explicitly incorporate firm-level heterogeneity observed in more than one dimension and study its interaction with aggregate macroeconomic fluctuations. Our methodology employs dimensionality reduction techniques for tensor data objects to approximate the joint distribution of firm-level characteristics. More broadly, our framework can be used for assessing predictions from structural models that account for micro-level heterogeneity observed on multiple dimensions. Leveraging firm-level data from the Compustat database, we use the FunVAR model to analyze the propagation of total factor productivity (TFP) shocks, examining their impact on both macroeconomic aggregates and the cross-sectional distribution of capital and labor across firms. 

\vspace{0.5cm}
\emph{J.E.L Classification Code: C32; E32 }

\bigskip
\bigskip
\bigskip
\bigskip
\end{abstract}
\clearpage

\section{Introduction }

The study of household and firm heterogeneity plays a crucial role in understanding macroeconomic fluctuations. Extensive research has recently focused on household heterogeneity and its implications for aggregate fluctuations within both fully structural and semi-structural frameworks \citep{aiyagari1994uninsured,kaplan2018monetary,bayer2019precautionary,bilbiie2023inequality,schorchang2022effects}. Analogously, heterogeneous firm models often suggest that firm heterogeneity offers additional nuanced insights for understanding macroeconomic fluctuations \citep{winberry2018,koby2020aggregation,winberry2021lumpy, ottonello2020financial}.  However, most contributions related to firm heterogeneity and its effects for macroeconomic fluctuations remain fully structural. Most of the times, these fully structural models are calibrated or estimated using only macroeconomic data.\footnote{An exception is \citet{liuplagbor} which devise an approach for estimating heterogeneous agents models exploiting both micro data and macroeconomic aggregates, but only apply this algorithm on simulated firm data.} Since firm heterogeneity is often neglected in semi-structural models, which are typically used to evaluate the effect of macroeconomic shocks on aggregate fluctuations, it is crucial to develop semi-structural models that can incorporate this heterogeneity. At the same time, these models should be flexible enough to deviate from strict theoretical predictions to allow for an effective empirical validation of theories using data.

\begin{figure}[H]
    \centering
    \begin{subfigure}{0.67\textwidth}
        \centering
        \includegraphics[width=\textwidth]{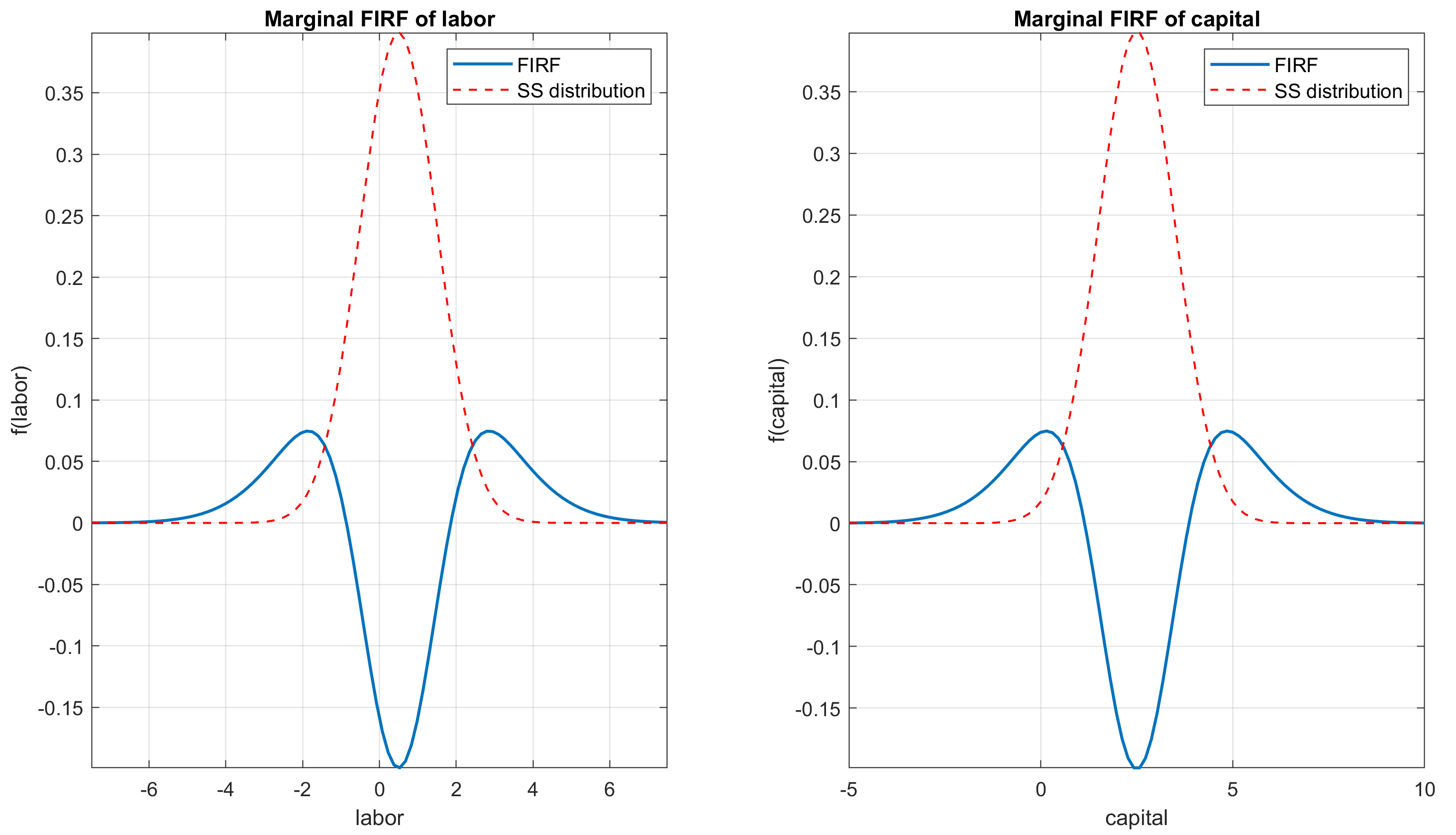}
        \label{fig:left_image}
    \end{subfigure}
    
    \vskip\baselineskip
    \begin{subfigure}{0.48\textwidth}
        \centering
        \includegraphics[width=\textwidth]{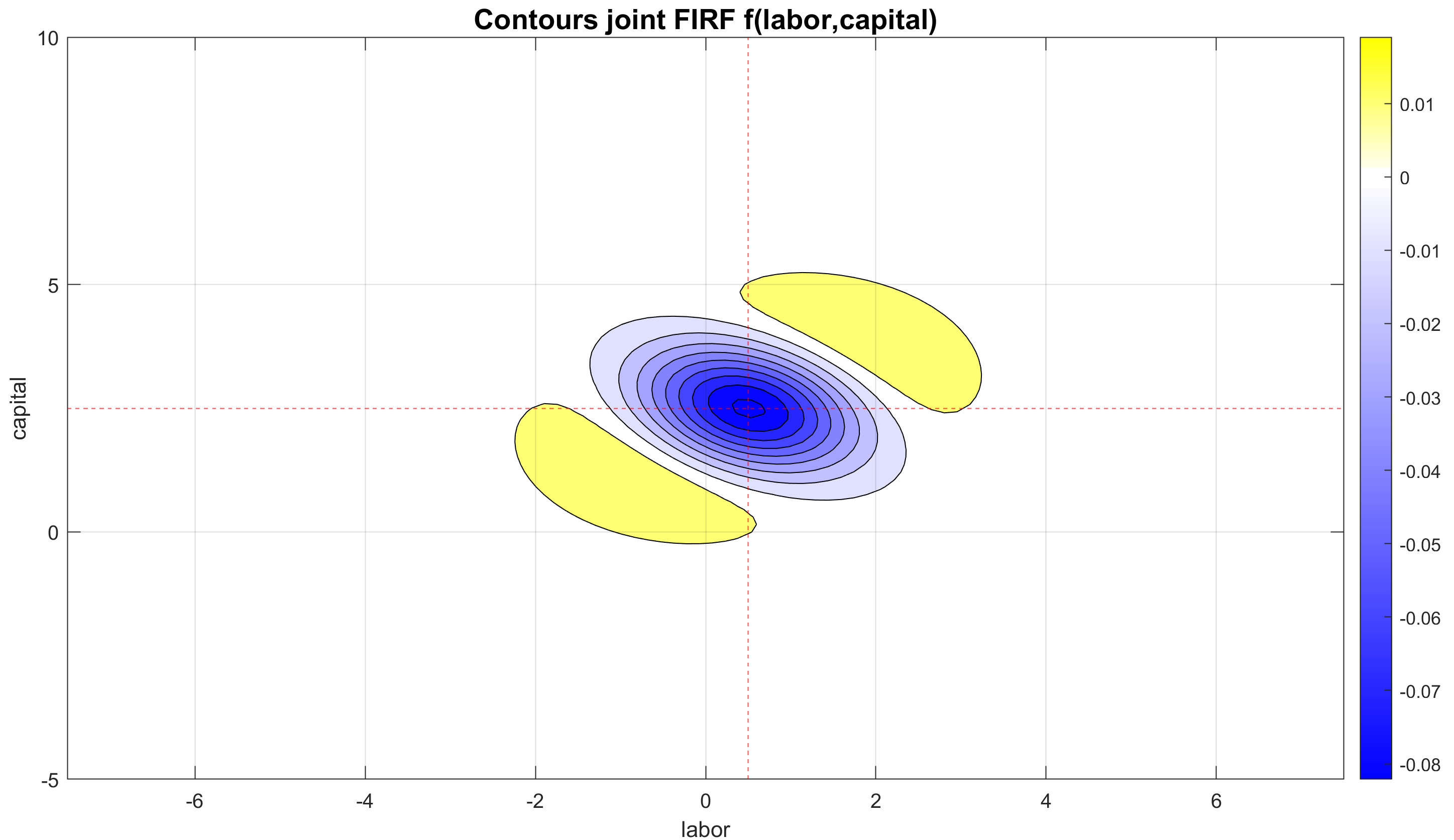}
        \label{fig:right_image1}
    \end{subfigure}
    \hfill
    \begin{subfigure}{0.48\textwidth}
        \centering
        \includegraphics[width=\textwidth]{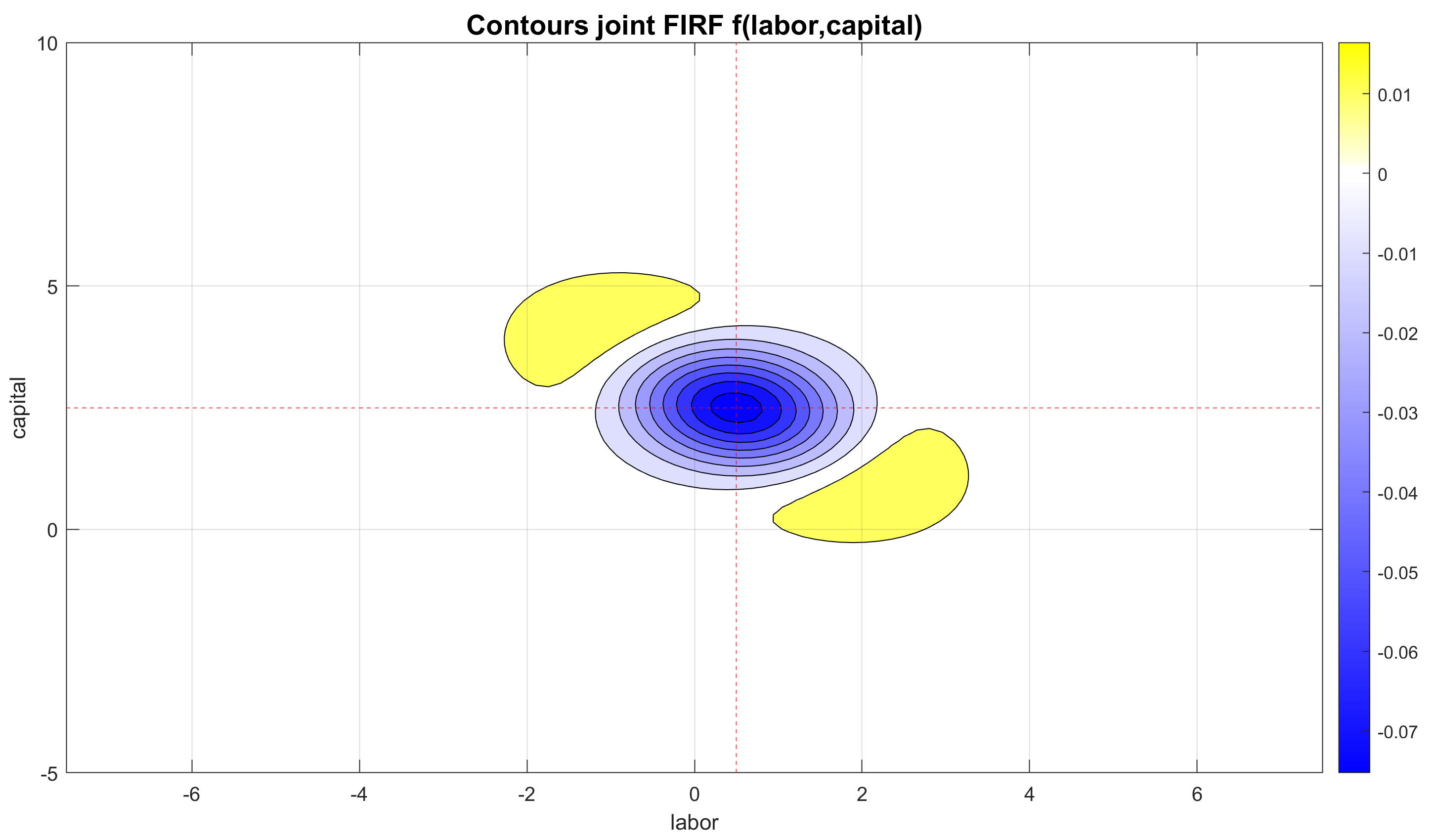}
        \label{fig:right_image2}
    \end{subfigure}
    \caption{Marginal and joint functional IRF of labor and capital}
    \label{fig:combined_images}
    \vspace*{-0.15cm}
    \parbox{1\textwidth}{\footnotesize{Notes: The figure shows the functional IRFs of the marginal labor and capital distributions to a shock (above). Below we report the contours from two different  bivariate functional IRFs both equally compatible with the changes in the marginal distributions above. In the figures below the red dashed line reports the steady state mean values.}}
\end{figure}

In this paper we propose a semi-structural model utilizing a Functional Augmented Vector Autoregression (FunVAR) to account for firm heterogeneity observed in more than one dimension and  assess its interaction with the macroeconomic aggregates. Our approach is inspired by recent advances in the estimation of semi-structural models, to assess the role of heterogeneity in understanding macroeconomic fluctuations and to test economic theories with heterogeneous agents \citep{schorchang2021heterogeneity} using both micro-level and aggregate macroeconomic data. Differently from previous contributions, our approach is designed for cases in which one wants to account for micro-level heterogeneity observed across multiple dimensions. In particular,  in the FunVAR model, we focus directly on modeling the joint distribution of firm-level characteristics, rather than marginal distributions, capturing the idea that structural aggregate shocks impact the full joint distribution. This joint modeling is essential for qualitatively assessing predictions from structural heterogeneous agent models, exploiting observations for more than one micro-level variable. Suppose for example that we have two structural heterogeneous firm models that differ fundamentally in terms of how firm's reallocate labor and capital after the realization of a shock.  In one structural model, after the aggregate shock, some firms with intermediate labor and capital levels expand both labor and capital while other firms  reduce both inputs. In the other structural model, some firms with intermediate labor and capital increase capital while reducing labor, while others do the opposite. Consider the case in which we specify a FunVAR in terms of the marginal distributions of labor and capital to understand whether the data favor one scenario over the other. Figure \ref{fig:combined_images}, upper panel, shows (in blue) the changes in mass of the marginal distribution functions after the realization of the shock, relative to the steady-state value, alongside the steady-state distribution itself (in red). 

After the shock, the mass of firms with both low labor and low capital endowments increases, as does the mass of firms with high labor and capital endowments. However, the marginals alone do not indicate how capital and labor are adjusting relative to each other. Importantly, the same changes in the marginal distributions are compatible with both structural models. In the lower panel, the figure shows two contour plots of the change in the mass of the bivariate density of labor and capital with respect to the bivariate steady-state distribution which are both consistent with the same changes in the marginals. In the scenario on the left, some firms with intermediate levels of capital and labor expand both inputs while others reduce both. In the scenario on the right, some firms specialize by increasing capital while reducing labor, while others specialize by increasing labor and reducing capital. Therefore, while challenging, modeling the joint distribution rather than just the marginal distributions of the observed micro variables is essential for qualitatively evaluating the effects of economic shocks on the micro-level distributions and understanding whether the data favor a certain scenario, supported for example by a specific structural model, over another. 

In this paper, we address this challenge by leveraging dimensionality reduction techniques for tensor data objects to non-parametrically approximate the multidimensional distribution of firm-level characteristics. These methods have been widely applied in feature extraction and pattern recognition tasks, particularly in contexts where data, such as 2-D/3-D images and video sequences, are naturally represented as tensors. We use these methods to non-parametrically approximate an unconstrained transformation of our multidimensional distribution function, that in our application is the firm-level joint labor and capital distribution.   These techniques make it easier to handle complex, high-dimensional dependencies compared to splines, Bernstein polynomials, and other basis function approximations. They also offer better scalability in multidimensional environments, where such basis functions can become computationally intensive and impractical. The finite dimensional approximation of the Functional VAR model with this approach leads to factor augmented VAR model. We cast the factor augmented VAR model in a state space form and estimate the model in two steps.

Another contribution of the paper is more specifically related to the analysis of firm heterogeneity and aggregate fluctuations. As anticipated above, structural heterogeneous firms model are often calibrated or estimated using macro data only. Similarly, semi structural models as VARs, which are used to study the propagation of macroeconomic shocks (see \citet{RAMEY201671} for a general review) typically neglect firm level heterogeneity and are often estimated without incorporation firm-level micro data.\footnote{An exception is \citet{lenza2024we}, that recently study the role of heterogeneity in the revenues of individual firms for the transmission of a business cycle shock in the euro area using the framework of \citet{schorchang2021heterogeneity}.} In the empirical analysis, we use the FunVAR to evaluate the effects of aggregate TFP shocks both on macroeconomic aggregates and on the joint distribution of firms-level labor and capital. We employ the Compustat database, exploiting both the Quarterly and the Annual datasets to recover the cross sectional distribution of firm-level capital and labor for the US companies from 1984-Q4 to 2019-Q4.
We find that after a TFP shock, firms generally increase their capital levels, even if labor remains below its steady state for some of them. The shock causes a shift where fewer firms combine high labor with low capital, while more firms accumulate both capital and labor above steady-state levels. 

The paper is structured as follows. Section \ref{sec:model} introduces the FunVAR model. Section \ref{sec:est_inf} discusses estimation and inference. Section \ref{sec:simul} tests the FunVAR using simulated data from an heterogeneous agents model. Section \ref{sec:emp_appl} studies the distributional effects of TFP shocks in the US. Section \ref{sec:concl} summarizes and concludes.

\section{Model}\label{sec:model}
We assume that we observe a vector of macroeconomic variables $\boldsymbol{y}_t$ and repeated cross sections of firms level characteristics $\boldsymbol{x}_{it}$ for $t=1,\ldots, T$ periods and $i =1, \ldots, N_t^{cross}$ firms. We also assume that the cross-sectional observations are drawn from a time-varying multivariate distribution with density $f_t(\boldsymbol{x})$. To model the dynamic interaction between the distribution function of firms-level characteristics and the aggregate macroeconomic time series, we consider the following function augmented VAR model

\begin{equation}\label{fvar}
        \boldsymbol{y_t} = \boldsymbol{c_y} + \sum_{s=1}^p \boldsymbol{B}_{l,yy}\boldsymbol{y_{t-s}} + \sum_{s=1}^p \int \boldsymbol{B_{s,yl}}(\boldsymbol{x})l_{t-s}(\boldsymbol{x}) d\boldsymbol{x} + \boldsymbol{u_{y,t}} \hspace{0.1cm},
\end{equation}
\begin{equation}\label{fvar2}
        l_{t}(\boldsymbol{x}) = c_l(\boldsymbol{x}) + \sum_{s=1}^p \boldsymbol{B}_{s,ly}(\boldsymbol{x})\boldsymbol{y}_{t-s} + \sum_{s=1}^P\int B_{ll}(\boldsymbol{x},\boldsymbol{x}')l_{t-s}(\boldsymbol{x}')d\boldsymbol{x}' + u_{l,t}(\boldsymbol{x}) \hspace{0.1cm}.     
\end{equation}
The macroeconomic aggregates are stored in the $n_y \times 1$ column vector $\boldsymbol{y_t}$ while $l_{t}(\boldsymbol{x})$ is defined to be the Centered-Log Ratio (CLR) transform of the multivariate density function of the vector of firms-level characteristics. For example, assuming that in each period we observe firms' specific labor and capital endowments, we have $\boldsymbol{x}= [x_1, x_2]'$ where $x_1$ stands for the labor input while $x_2$ for the capital input. The function $l_{t}(\boldsymbol{x})^{obs}$ summarizes the information about the distribution of the firm-level capital and labor at each time $t$. In particular the  CLR transformation of the distribution function is given by
\begin{equation}
    l_t(\boldsymbol{x}) : =\text{CLR}(f_t(x_1, x_2)) = \log(f_t(x_1, x_2)) - \frac{1}{|\Omega|} \int_{\Omega} \log(f_t(x_1, x_2)) \, dx_1 dx_2 \hspace{0.1cm}.  
\end{equation}
This transformation greatly simplifies the econometric analysis of the time variation of the multivariate distribution function as it maps a density function, that needs both to integrate to one and to obey non-negativity constraints, to an unconstrained real-valued space.\footnote{The CLR transformation, traditionally used in the context of compositional data, has been used for modelling distribution functions by  \citet{menafoglio}. For an extensive discussion on the use of the CLR transformation in the context of distribution functions we refer to \citet{petersenkokosca2021}.} We assume that $l_t(\boldsymbol{x})$ that we can observe or estimate on a grid is a noisy realization of the CLR transformation of the true multivariate density function, namely

\begin{equation}\label{eq_err} 
   l_t(\boldsymbol{x})^{\text{obs}} = l_t(\boldsymbol{x}) + \varepsilon_t, \quad 
\end{equation}
where $\varepsilon_t$ is a noise with $\E[\varepsilon_t] = 0 $ and $\E[\varepsilon_t^2] = \sigma^2$ and $l_t(\boldsymbol{x})$ is the true CLR transformation of the distribution function evaluated at the grid points. For example, in the bivariate labor and capital example we can denote \( \mathcal{X}_1 = \{x_{1,1}, x_{1,2}, \dots, x_{1,N_1}\} \) and \( \mathcal{X}_2 = \{x_{2,1}, x_{2,2}, \dots, x_{2,N_2}\} \) the sets of grid points for labor and capital, respectively, and define \( \boldsymbol{x} \in \mathcal{X}_1 \times \mathcal{X}_2 \), the Cartesian product of all pairs \( (x_1, x_2) \).  We can then estimate $f_t(x_1,x_2)$ from available data obtaining noisy observations of the true CLR transformation of the distribution function on $N^{grid} = N_1N_2$ grid points, where the noise is due to the density estimation error.
We assume that the true centered log-ratio transformation of the multivariate density function admits the following finite basis expansion 
\begin{equation}
    l_t(x_1,x_2) = \sum_{i=1}^{K} \beta_{i,t} h_i(x_1,x_2) \hspace{0.1cm},  
\end{equation}
which in a more general multidimensional setting can be written as
\begin{equation}\label{finite_approx}
        l_t(\boldsymbol{x}) = \sum_{i=1}^{K} \beta_{i,t} h_i(\boldsymbol{x}) \hspace{0.1cm}.  
\end{equation}
This expansion let us to rewrite the functional VAR model as a factor augmented VAR model for the aggregate macroeconomic variables $\boldsymbol{y}_t$ and the factors $\boldsymbol{\beta_t}$, that is\footnote{In the appendix \ref{sec_approx} we report the steps, which directly follow \citet{schorchang2021heterogeneity}, to derive the factor augmented representation of the functional VAR.}
\begin{equation}\label{favar}
\begin{aligned}
\begin{bmatrix}
    \boldsymbol{y}_t \\
    \boldsymbol{\beta}_{t}  \\
    \end{bmatrix} =  \boldsymbol{\Phi}_0 + 
    \boldsymbol{\Phi}_1
    \begin{bmatrix}
    \boldsymbol{y}_{t-1} \\
    \boldsymbol{\beta}_{t-1}  \\
    \end{bmatrix} + \ldots +  
     \boldsymbol{\Phi}_p
    \begin{bmatrix}
    \boldsymbol{y}_{t-p} \\
    \boldsymbol{\beta}_{t-p}  \\
    \end{bmatrix} +
    \begin{bmatrix}
    \boldsymbol{u}_{y,t}  \\
    \boldsymbol{\tilde{u}}_{l,t}   \\ 
    \end{bmatrix} \hspace{0.1cm}.
\end{aligned} 
\end{equation}
In this factor augmented VAR model, the dynamic behavior of the factors $\boldsymbol{\beta}_t$ is governing the time variation of firm-level characteristics over time. Macroeconomic shocks, such as aggregate TFP shocks, are driving the joint dynamics of the cross-sectional distribution of firm level characteristics and the macroeconomic aggregates through the vector $[\boldsymbol{\tilde{u}}_{y,t}' \boldsymbol{\tilde{u}}_{l,t}']$, that similarly to the standard Structural VAR framework, can be interpreted as linear combinations of the structural shocks. 

\section{Estimation and inference}\label{sec:est_inf}

The CLR transformation maps the multivariate density function $f_t(\boldsymbol{x})$ to the space \(L^2 \) of square-integrable real measurable functions, where statistical methods for unconstrained data can be applied. For any $t$ we compute the CLR transformation as follows: 
\begin{equation}
l(\boldsymbol{x})^{\text{obs}} = CLR(\hat{f}(\boldsymbol{x})) \hspace{0.1cm},
\end{equation}
where $\hat{f}(.)$ is a kernel density estimate of the distribution function on the set of grid points $N^{grid}$. For example, in the bivariate labor and capital example, it is obtained as 
\begin{equation}  
\hat{f}(x_1, x_2) = \frac{1}{N_1 N_2 h_1 h_2} \sum_{i=1}^{N_1} \sum_{j=1}^{N_2} K\left( \frac{x_1 - x_{1,i}}{h_1}, \frac{x_2 - x_{2,j}}{h_2} \right) \hspace{0.1cm}.
\end{equation}
Considering all the points in the grid, namely $\{ l_t(x_{1,i}, x_{2,j})^{\text{obs}} : i = 1, \dots, N_1, \, j = 1, \dots, N_2 \}$   and storing them in the matrix, $\boldsymbol{L}_t$, our observation equation becomes
\begin{equation}
\boldsymbol{l}_t^{obs} : = vec(\boldsymbol{L}_t) = \boldsymbol{H}\boldsymbol{\beta}_t + \boldsymbol{\varepsilon}_t    \hspace{0.1cm},
\end{equation}
where  $vec(\boldsymbol{L}_t)$ is the $N^{grid} \times 1$ vector of observed values for the CLR transformation of the bivariate distribution function, $\boldsymbol{H}$ is the $N^{grid} \times K$ matrix of loadings, while $\boldsymbol{\beta}_t$ is the $K \times 1$ vector of scores and $\boldsymbol{\varepsilon}_t $ is the vector of noises. To estimate the model, we proceed in two steps following the approach of 
\citet{DOZGIANRIECH} for dynamic factor models. First we estimate the loadings $\boldsymbol{H}$, needed for the basis expansion approximation. Once we have obtained an estimate of the loadings, in the second step, we perform Bayesian posterior inference on the parameters of the VAR model (\ref{favar}) i.e. $(\boldsymbol{\Phi},\boldsymbol{\Sigma})$ and the latent factors $\boldsymbol{\beta}_{1:T}$. In the next sections, we describe three approaches for estimating the loadings in $\boldsymbol{H}$, under a different set of assumptions concerning the approximation of the true CLR trasformed multidimensional distribution function in (\ref{finite_approx}). We then detail the estimation of the factor augmented VAR model approximation of the FunVAR model.  

 
 \subsection{Unfolding and approximation by principal component analysis}\label{sec:unfolding}
One approach for estimating the loadings $\boldsymbol{H}$ needed for the approximation of the multidimensional distribution function consists in vectorizating the matrix of observable $\boldsymbol{L}_t = CLR (\boldsymbol{\hat{F}_t})$, where $\boldsymbol{\hat{F}_t}$ is the matrix containing the value of $\hat{f}(x_1, x_2)$ when evaluated on the chosen grid, and then performing principal component analysis on the $T$ vectors $\boldsymbol{l}_t^{obs}$, each of dimension $N^{grid} \times 1$. This practice is known as flattening or unfolding in the context of tensor analysis \citep{tensor_rev}. In particular, going back to the labor and capital example, we can define the tensor object $\mathcal{L} \in \mathcal{R}^{N_1 \times N_2 \times T}$, which is storing the values of the CLR transformation of the multivariate density function on the finite grid for labor and capital, for all the time periods in the sample. First, we unfold the tensor obtaining the $N_1N_2 \times T$ matrix $\boldsymbol{\tilde{L}}$. Then, we perform principal component analysis on $\boldsymbol{\tilde{L}}$ since this matrix is stacking on its columns the $N_1N_2$ dimensional column vectors $\boldsymbol{l}_t^{obs}$ for $t =1, \ldots, T$. Doing so we aim estimating the $N_1N_2 \times K$ loading matrix $\boldsymbol{H}$ minimizing the reconstructing error, that is
\begin{equation}
\min_{\boldsymbol{H},\boldsymbol{\beta}_t} \frac{1}{T} \sum_{t=1}^{T} \left\| vec(\boldsymbol{L}_t) - \boldsymbol{H} \boldsymbol{\beta}_t  \right\|^2 \hspace{0.1cm}.
\end{equation}
To estimate $\boldsymbol{H}$ we apply the SVD decomposition to  
 the unfolded matrix $\boldsymbol{\tilde{L}}$, that is
\[
\boldsymbol{\tilde{L}} = \boldsymbol{U S V^T},
\]
where $\boldsymbol{U}$ and $\boldsymbol{V}$ are orthogonal matrices, and $\boldsymbol{S}$ is a diagonal matrix containing the singular values. More specifically, we select \( K \) columns from \( \boldsymbol{V} \), which represent the eigenbasis associated with the \( K \) largest eigenvalues. Note that, in this approach, PCA on the flattened data identifies the top K modes of variation without distinguishing between the individual dimensions.  Instead, it treats all dimensions as a single combined dimension, ignoring the multi-dimensional structure and the interactions specific to each dimension. The approach can be naturally extended to the case in which we have more than two dimensions, for example when $N_C$ is the number of micro-level variables, we unfold the $N_C +1$ dimensional tensor  $\mathcal{L} \in \mathcal{R}^{N_1 \times N_2 \times \ldots \times N_C \times T}$ into a $(\prod_{i=1}^NN_i) \times T$ matrix and then perform principal component analysis for identifying the top K modes of variation. Note also that, in general, this approach requires the estimation of $N^{grid} \times K$ loadings in the $\boldsymbol{H}$ matrix, which in small samples can be challenging. In the next sections we consider two approaches for reducing the number of parameters to be estimated, assuming specific basis expansions for the CLR transformation of the multivariate density function.

\subsection{Approximation by multilinear principal component analysis}
We now consider an alternative approach to the practice of performing principal component analysis on the unfolded data in the previous section. This approach considerably reduces the number of loadings to be estimated and explicitly leverages the multi-dimensional structure of the data, applying dimensionality reduction to each mode separately. This methodology aligns with the Tucker decomposition framework \citep{tucker1966some}, which allows for the decomposition of a tensor into mode-specific factors and a core tensor. Specifically, going back to the example on the approximation of the bivariate labor and capital distribution, we assume that the true centered log ratio transformation of the density function can be expanded as
\begin{equation}\label{kronexpansio_basis}
    l_t(x_1,x_2) = \sum_{i=1}^{K_1}\sum_{j=1}^{K_2}  \beta_{ij,t} h_i(x_1) h_j(x_2)
\end{equation}
Note that this expansion is a particular case of the more general basis expansion in (\ref{finite_approx}).
The representation is bilinear, meaning it is expressed as a product of components that vary across two separate dimensions, that is labor $x_1$ and capital $x_2$. This structure assumes that the function lies in a lower-dimensional subspace (rank $K_1 \times K_2$) of the full space spanned by the basis functions. 
In practice this implies a dimensionality reduction, as it approximates the full function by focusing on the most important modes of variation in the two separate dimensions (controlled by 
$K_1$ and $K_2$). The expansion can be written using the Kronecker product as
\begin{equation}\label{kronexpansion}
    l_t(x_1,x_2) =  (\boldsymbol{h}(x_2) \otimes \boldsymbol{h}(x_1))' \boldsymbol{\beta}_t \hspace{0.1cm}.
\end{equation}
where $\boldsymbol{h}(x_1) \otimes \boldsymbol{h}(x_2)$ is of dimension $ K \times1  $ where now $K =  K_1 K_2$, while $\boldsymbol{\beta}_t$ is of dimension  $K \times 1$:  

\[
\boldsymbol{h}(x_1) = \begin{bmatrix}
h_1(x_1) \\
h_2(x_1) \\
\vdots \\
h_{K_1}(x_1)
\end{bmatrix}, \quad
\boldsymbol{h}(x_2) = \begin{bmatrix}
h_1(x_2) \\
h_2(x_2) \\
\vdots \\
h_{K_2}(x_2)
\end{bmatrix}, \quad
\boldsymbol{\beta}_t = \begin{bmatrix}
\beta_{11,t} \\
\beta_{21,t} \\
\vdots \\
\beta_{K_11,t} \\
\beta_{12,t} \\
\vdots \\
\beta_{K_1K_2,t}
\end{bmatrix} \hspace{0.1cm}.
\]
Therefore, in terms of the observable, (\ref{eq_err}) and (\ref{kronexpansion}), translate in 
the following bilinear form 
\begin{equation}\label{bilinearform}
\boldsymbol{L}_t = \boldsymbol{H}_1\boldsymbol{B_t} \boldsymbol{H}_2'  + \boldsymbol{E_t} \hspace{0.1cm},
\end{equation}
where $\boldsymbol{L}_t$ is $N_1 \times N_2$ , $\boldsymbol{H}_1$ and $\boldsymbol{H}_2$ are the basis functions evaluated on the grid of values for labor $x_1$ and capital $x_2$, respectively of dimension \(N_1 \times K_1\)  and \(N_2 \times K_2\). $\boldsymbol{B}_t$  is the \(K_1 \times K_2\) matrix of  factors at time \(t\) while $\boldsymbol{E}_t$ is the $N_1 \times N_2$ matrix of noises. Vectorizing equation (\ref{bilinearform}) and exploiting the properties of the Kronecker product we get
\begin{equation}
\boldsymbol{l}_t^{obs} =  (\boldsymbol{H}_2 \otimes \boldsymbol{H}_1) \boldsymbol{\beta_t} + \boldsymbol{\varepsilon_t} \hspace{0.1cm},
\end{equation}
where  $\boldsymbol{l}_t^{obs}$ is the $N^{grid}\times 1$ vector of observable with $N^{grid} = N_1  N_2$, $\boldsymbol{\beta}_t = vec(\boldsymbol{B}_t)$ and $\varepsilon_t = vec(\boldsymbol{E}_t)$. Note that the number of loadings is now equal to $N_1 K_1 + N_2 K_2$, while in the previous approach based on principal component analysis on the vectorized data we had $N_1N_2 \times K$ loadings. We apply bilinear principal component analysis to estimate the loadings in $\boldsymbol{H} = (\boldsymbol{H}_2 \otimes \boldsymbol{H}_1)$.  Estimation by bilinear principal component (see \citet{ye2004generalized}) seeks to minimize:
\begin{equation}
\min_{\boldsymbol{H}_1, \boldsymbol{H}_2, \{\boldsymbol{B}_t\}} \frac{1}{T} \sum_{t=1}^{T} \left\| \boldsymbol{L}_t - \boldsymbol{H}_1 \boldsymbol{B}_t \boldsymbol{H}_2' \right\|^2  \hspace{0.1cm}.
\end{equation}
Theorem 1 in \citet{biometrika_bilinear} let us re-frame the problem in terms of the population expected Frobenius norm $\E \{||\boldsymbol{L} - \boldsymbol{H}_1\boldsymbol{B}\boldsymbol{H}_2' ||^2 \}$. In particular, the minimizers $\boldsymbol{H}_1 \in \mathcal{O}_{K_1,\tilde{K}_1}$ and $\boldsymbol{H}_1 \in \mathcal{O}_{K_2,\tilde{K}_2}$ will be equal to the maximizers of the following problem:
\begin{equation}
\max_{\boldsymbol{H}_1, \boldsymbol{H}_2}   \E \{||\boldsymbol{H}_1'\boldsymbol{L}\boldsymbol{H}_2 ||^2 \}  \hspace{0.1cm}.
\end{equation}
We use an iterative approach to estimate \(\boldsymbol{H}_1\) and \(\boldsymbol{H}_2\). Starting with initial random matrices \(\boldsymbol{H}_1^{(0)}\), \(\boldsymbol{H}_2^{(0)}\) we iterate the following step
\begin{enumerate}
    \item  For fixed \(\boldsymbol{H}_2^{(k)}\)  update $(\boldsymbol{H}_1^{(k+1)})$ by maximizing 
     \begin{equation}
    \boldsymbol{H}^{(k+1)}_1 = \text{argmax}_{\boldsymbol{H}_1} T^{-1}\sum_{t=1}^{T} \left\|\boldsymbol{H}_1 \boldsymbol{L}_t \boldsymbol{H}_2'^{(k)} \right\|^2  \hspace{0.1cm}.
    \end{equation}   
    \item For fixed \(\boldsymbol{H}_1^{(k+1)}\)  update $(\boldsymbol{H}_2^{(k+1)})$ by maximizing
    \begin{equation}
    \boldsymbol{H}_2^{(k+1)} = \text{argmax}_{\boldsymbol{H}_2} T^{-1}\sum_{t=1}^{T} \left\|\boldsymbol{H}_1^{(k+1)} \boldsymbol{L}_t \boldsymbol{H}_2' \right\|^2  \hspace{0.1cm}.
    \end{equation}
\end{enumerate}
Both maximization problems are standard eigenvalue problems, which can be formulated in terms of Singular Value Decompositions (SVD). As the algorithm may find only a local maximum, multiple random initial values are considered  to ensure that the global maximum is found.  Iterations stop when the change in the objective function is found to be lower than a predefined small $\epsilon$, indicating convergence. \citet{biometrika_bilinear} develop the asymptotic theory for this type of order-two multilinear principal component analysis, addressing both asymptotic efficiency and the distributions of the principal components and associated projections. Their work provides comprehensive details on convergence rates and efficiency, which we refer to for further details. Additionally, they propose a method for selecting the dimensionality parameters 
$K_1$ and $K_2$ based on a test concerning the explained proportion of total variance. This test determines whether the variance explained by the selected dimensions exceeds a predefined threshold, to which we also refer for guidance on dimensionality selection. 
Because principal component analysis on unfolded data requires the estimation of many parameters, multilinear principal component analysis is expected to outperform conventional principal component analysis, especially when the sample size is small to moderate. However, principal component analysis in unfolded data may provide better approximations when these interactions are weak or noisy, or the basis expansion assumed in the Tucker decomposition is too restrictive. 

In the multilinear principal component analysis, it is also natural to handle more than two dimensions. For example, in the $N_C$ dimensional case the basis expansion of the true CLR transformation of the distribution function just becomes 
\begin{equation}\label{kronecker_expansion}
    l_t(x_1, x_2, \dots, x_{N_C}) = (\boldsymbol{h}(x_{N_C})  \otimes \dots \otimes \boldsymbol{h}(x_2) \otimes \boldsymbol{h}(x_{1}))' \boldsymbol{\beta}_t \hspace{0.1cm},
\end{equation}
where $(\boldsymbol{h}(x_{N_C})  \otimes \dots \otimes \boldsymbol{h}(x_2) \otimes \boldsymbol{h}(x_{1}))$ is of dimension $\prod_{i=1}^{N_C}{K_i}  \times 1$. In terms of the observables we have
\begin{equation}\label{tucker_decomposition}
    \mathcal{L}_t = \mathcal{B}_t \times_1 \boldsymbol{H}_1 \times_2 \boldsymbol{H}_2 \times_3 \boldsymbol{H}_3 \dots \times_{N_C} \boldsymbol{H}_{N_C} + \mathcal{E}_t \hspace{0.1cm},
\end{equation}
where the notation $\times_n$ refers to the mode-n product of a tensor and a matrix and $ \mathcal{B}_t$ is the core tensor of the Tucker decomposition. The iterative procedure described above is then just extended to estimate all the matrices with the loadings $\boldsymbol{H}_1, \ldots, \boldsymbol{H}_{N_C}$, each of dimension $N_i \times K_i$ for $i =1, \ldots, N_C$.


\subsubsection{CANDECOMP/PARAFAC decomposition}
The multilinear decomposition is particularly well-suited for datasets with interactions across multiple modes, since it permits to perform dimensionality reduction in a flexible manner.
An alternative to the multilinear decomposition is the CP decomposition (CANDECOMP/PARAFAC) \citep{Carroll1970, Harshman1970}. In this decomposition, the rank is shared across all modes, which means that the same number of components is used for each dimension of the tensor. In particular, going back to the bivariate labor and capital example, it is assumed that:
\begin{equation}
    l(x_1,x_2) =  \sum_{k=1}^K \beta_{t,k} h_{k}^{(1)}(x_1) h_{k}^{(2)}(x_2) \hspace{0.1cm}.
\end{equation}
Note that in this case the number of loadings to be estimated becomes $K(N_1 + N_2)$. The loadings can be estimated by Alternating Least Squares (ALS) \citep{Carroll1970,Harshman1970}. \citet{babii2024tensor} have recently shown how to estimate the loadings of the CP decomposition by iterative principal components on the unfolded tensor along each of its dimensions. They label this estimator Tensor-PCA and derive its asymptotic properties. The CP decomposition framework also naturally extends to more than two dimensional settings, developing the expansion: 
\begin{equation}
    l(\boldsymbol{x}) =  \sum_{k=1}^K \beta_{t,k} \prod_{i=1}^{N_C} h_{k}^{(i)}(x_i) \hspace{0.1cm}.
\end{equation}
In terms of the observables we have:
\begin{equation}\label{CP_decomposition}
    \mathcal{L}_t = \sum_{k=1}^K \beta_{t, k} \bigotimes_{i=1}^{N_C} \boldsymbol{h}_k^{(i)} + \mathcal{E}_t,
\end{equation}
where $\bigotimes_{i=1}^{N_C} \boldsymbol{h}_k^{(i)}$ denotes the outer product of vectors  $\boldsymbol{h}_k^{(i)}$  across each dimension $i = 1, 2, \dots, N_C$. In general, the CP decomposition is less flexible when different modes of the data exhibit varying complexity or correlation structures.  In contrast, the multilinear decomposition allows for different ranks in each mode, enabling more precise control over dimensionality reduction in each dimension.

\subsection{Factor augmented VAR}
Conditional on the estimate of the loadings $\boldsymbol{H}$ in the first step, obtained through one of the three alternative approaches proposed above, we perform Bayesian inference on the parameters of the factor augmented VAR model: 
\begin{equation}\label{obsW}
    \boldsymbol{w}_t = \boldsymbol{\Phi}\boldsymbol{x}_t +  \boldsymbol{u}_t \hspace{0.1cm} , \hspace{1cm} \boldsymbol{u}_t \sim \mathcal{N}(0, \boldsymbol{\Sigma}) \hspace{0.1cm},
\end{equation}
\begin{equation}\label{obsH}
\boldsymbol{l}_t^{obs} = \boldsymbol{H}\boldsymbol{\beta}_t + \boldsymbol{\varepsilon}_t \hspace{0.1cm}, \hspace{1cm} \boldsymbol{\varepsilon}_t \sim  \mathcal{N}(0, \boldsymbol{I}_{N^{grid}},\sigma^2)   \hspace{0.1cm},
\end{equation}
where  $\boldsymbol{w}_t = [\boldsymbol{y}_t,\boldsymbol{\beta}_t']' $ 
    and $\boldsymbol{x}_t = [1,\boldsymbol{w}_{t-1}', \ldots, \boldsymbol{w}_{t-p}']'$. We cast the model in state space form and treat $\boldsymbol{\beta_{1:T}}$ as latent stochastic states. We need to specify a prior for distribution for $(\boldsymbol{\Phi},\boldsymbol{\Sigma})$ and for the noise variance $\sigma^2$. A common approach is to use the standard Normal-Inverse Wishart conjugate prior for $(\boldsymbol{\Phi},\boldsymbol{\Sigma})$, which is computationally attractive in large-dimensional settings \citep{CARRIERO2009400,bamburagiannoneriechlin}. However, this imposes symmetric shrinkage across all equations in the VAR, which can be restrictive. To address this issue, we utilize the asymmetric conjugate prior proposed by \citet{chan2022asymmetric}, which permits to estimate the VAR parameters in high-dimensional settings while maintaining computational feasibility and enabling asymmetric shrinkage across equations. For $\sigma^2$,  instead, we specify a standard independent Inverse Gamma prior. We combine these prior distributions  with the likelihood implied by (\ref{obsH}) and (\ref{obsW}), obtaining the following posterior distribution for ($\boldsymbol{\beta}_{1:T},\sigma^2,\boldsymbol{\Phi},\boldsymbol{\Sigma}$), 
\begin{equation}\label{posterior}
p(\boldsymbol{\beta}_{1:T},\sigma^2,\boldsymbol{\Phi},\boldsymbol{\Sigma}|\boldsymbol{y}_{1:T},\boldsymbol{l}_{1:T}) \propto p(\boldsymbol{y}_{1:T}|\boldsymbol{\beta}_{1:T},\boldsymbol{\Phi},\boldsymbol{\Sigma})p(\boldsymbol{l}_{1:T}| \boldsymbol{\beta}_{1:T}, \sigma^2) p(\boldsymbol{\Phi},\boldsymbol{\Sigma})p(\sigma^2)  \hspace{0.1cm}.
\end{equation}
In order to perform inference on (\ref{posterior}), we device a \textit{Gibbs sampler} 
 that iteratively draws from the following conditional posterior distributions
\begin{enumerate}
    \item Draw from $p(\sigma^2|.)$.   
    
    \item Draw from $p(\boldsymbol{\Sigma}|.)$.
    \item Draw from $p(\boldsymbol{\Phi_0} , \boldsymbol{\Phi_1}, \ldots, \boldsymbol{\Phi_p}|.)$. 
    \item Draw from $p(\boldsymbol{\beta}_{1:T}|.)$.
\end{enumerate}
Step 1 to 3 of the Gibbs Sampler are standard. In order to draw from the conditional distribution of the latent states in step 4 of the Gibbs Sampler, we exploit the linearity of the system, and sample the vector of the latent states jointly in a single step from $ t = 1, \ldots, T$. This approach operationally borrows from  \citet{kauffman} and \citet{CHAN2023105468}. More precisely, we define: 
\begin{equation}
    \boldsymbol{w} = \boldsymbol{S}_o\boldsymbol{y} + \boldsymbol{S}\boldsymbol{b} \hspace{0.1cm},
\end{equation}
where $\boldsymbol{w} = ([\boldsymbol{y}_1' ,\boldsymbol{\beta}_{1}'],\ldots  ([\boldsymbol{y}_T' ,\boldsymbol{\beta}_{T}'])'= (\boldsymbol{w}_1',\ldots , \boldsymbol{w}_T'])' $ is a $T(n_y + K) \times 1$ vector that stores the macro aggregates together with the latent factors  from the cross sectional distribution, while $\boldsymbol{y} = (\boldsymbol{y}_1', \ldots, \boldsymbol{y}_T' )'$ and $\boldsymbol{b} = (\boldsymbol{\beta}_{1}', \ldots, \boldsymbol{\beta}_{T}')' $ respectively store the macroeconomic aggregates and the latent states. The selection matrices $\boldsymbol{S}_o$ and $\boldsymbol{S}$ are given by $\boldsymbol{S}_o = \boldsymbol{I}_T \otimes \boldsymbol{S}_{oo}$ and $\boldsymbol{S} = \boldsymbol{I}_T \otimes \boldsymbol{S}_{ff}$  with $\boldsymbol{S}_{oo} = [\boldsymbol{I}_{n_y};\boldsymbol{0}_{K\times n_y} ]$  and  $\boldsymbol{S}_{ff} = [\boldsymbol{0}_{n_y \times  K };\boldsymbol{I}_{K} ]$. Defining
\begin{equation}
\begin{aligned}
    \boldsymbol{c_{\Phi}} &= \begin{bmatrix}
        \boldsymbol{\Phi}_0 + \sum_{j=1}^{p} \boldsymbol{\Phi}_j \boldsymbol{w}_{1-j} \\
        \boldsymbol{\Phi}_0 + \sum_{j=2}^{p} \boldsymbol{\Phi}_j \boldsymbol{w}_{2-j} \\
        \vdots \\
        \boldsymbol{\Phi}_0 + \boldsymbol{\Phi}_p \boldsymbol{w}_0 \\
        \boldsymbol{\Phi}_0 \\
        \vdots \\
        \boldsymbol{\Phi}_0
    \end{bmatrix}, \quad
    \boldsymbol{{D}_{\Phi}} = \begin{bmatrix}
        \boldsymbol{I}& \boldsymbol{0}& \cdots & \boldsymbol{0} & \boldsymbol{0} \\
        -\boldsymbol{\Phi}_1 & \boldsymbol{I}& \boldsymbol{0} & \cdots & \boldsymbol{0} \\
        -\boldsymbol{\Phi}_2 & -\boldsymbol{\Phi}_1 & \boldsymbol{I} & \cdots & \boldsymbol{0} \\
        \vdots & \vdots & \ddots & \ddots & \vdots \\
        -\boldsymbol{\Phi}_p & \cdots & -\boldsymbol{\Phi}_3 & -\boldsymbol{\Phi}_2 & -\boldsymbol{\Phi}_1 \\
        \boldsymbol{0} & \cdots & \boldsymbol{0} & \boldsymbol{0} & \boldsymbol{I}
    \end{bmatrix} \hspace{0.1cm},
\end{aligned}
\end{equation}
and the products $\boldsymbol{G}_y = \boldsymbol{D_{\Phi}}\boldsymbol{S}_o$ and $\boldsymbol{G} =   \boldsymbol{D_{\Phi}}\boldsymbol{S}$,
we can rewrite the factor augmented VAR as: 
\begin{equation}\label{eq1}
    \boldsymbol{G}_o \boldsymbol{y} + \boldsymbol{G} \boldsymbol{b} =  \boldsymbol{c_{\Phi}} + \boldsymbol{u} \hspace{0.1cm}, \hspace{1cm } \boldsymbol{u} \sim \mathcal{N}(\boldsymbol{0}, \boldsymbol{I}_T\otimes \boldsymbol{\Sigma}) \hspace{0.1cm},
\end{equation}
and the observation equation in terms of $\boldsymbol{\beta}$ as follows: 
\begin{equation}\label{eq2}
    \boldsymbol{l} = \boldsymbol{M}\boldsymbol{b} + \boldsymbol{\epsilon} \hspace{0.1cm},  \hspace{1cm} \boldsymbol{\epsilon} \sim \mathcal{N}(\boldsymbol{0}, \sigma^2\boldsymbol{I}_{TN^{grid}})  \hspace{0.1cm},
\end{equation}
    where $\boldsymbol{l} = vec([\boldsymbol{l}_1, \ldots ,\boldsymbol{l}_T]) $ is $N^{grid}T \times 1$ vector of observable values of the CLR transformation of the multidimensional distribution function on the grid points, while $\boldsymbol{M} = \boldsymbol{I}_T \otimes \boldsymbol{H}$ is the matrix containing the loadings for the basis functions approximation of all cross-sectional distributions for $t=1,\ldots,T$.  Equations (\ref{eq1}) and (\ref{eq2}) imply that we can sample the latent states jointly from the conditional posterior distribution of $\boldsymbol{b}$ given by

\begin{equation}\label{condpostb}
  p(\boldsymbol{b}|.) \sim \mathcal{N}\left(\boldsymbol{\bar{K}}^{-1}\left(\frac{1}{\sigma^2}\boldsymbol{M}'\boldsymbol{l} + \boldsymbol{K} \boldsymbol{\mu} \right), \boldsymbol{\bar{K}}^{-1}  \right) \hspace{0.1cm}, 
\end{equation}
where \[\boldsymbol{\bar{K}} = \frac{1}{\sigma^2}\boldsymbol{M'M}  + \boldsymbol{G}'(\boldsymbol{I_T} \otimes \boldsymbol{\Sigma})^{-1}\boldsymbol{G} \hspace{0.1cm},\] with 
$\boldsymbol{\mu} = \boldsymbol{K}^{-1}\boldsymbol{G}'(\boldsymbol{I}_T \otimes \boldsymbol{\Sigma})(\boldsymbol{c_{\Phi}} - \boldsymbol{G}_{o}\boldsymbol{y} )$ and $\boldsymbol{K} = \boldsymbol{G}'(\boldsymbol{I}_T
\otimes \boldsymbol{\Sigma})^{-1}\boldsymbol{G}$. Note that the computational intensity of this step does not depend on the number of grid points on which the CLR transformation of the distribution function is evaluated $N^{grid}$, since the product matrices $\boldsymbol{M}'\boldsymbol{M}$ and $\boldsymbol{M'l}$ are respectively of dimension $TK \times TK$ and and $TK \times 1$ and are a function of $\boldsymbol{H}$ which is computed outside of the Gibbs Sampler. Computational complexity hinges instead on $K$, the number of basis used for the expansion of the CLR transformation of the distribution function. When the combination $TK$ is large, the latent scores can alternatively be drawn exploiting the standard Kalman filter and smoother step à la \citet{kc1994}. 
\subsubsection{Missing densities and mixed frequency}\label{mixedfrequency}
In applied work, the macroeconomic aggregate variables $\boldsymbol{y}$ and the distribution of firm's micro level characteristics $f(\boldsymbol{x})$ are often sampled at different frequencies. For example, the joint distribution on firm-level capital and labor for the US companies in the Compustat dataset is only available at the annual frequency, while most of the macroeconomic indicators are still available at the quarterly frequency. We adapt the standard factor augmented VAR framework, to allow for the missing densities. Our model becomes:
\begin{equation}\label{obsstatespace}
    l_{t}(\boldsymbol{x})^{obs} =  
\begin{cases}
  \boldsymbol{H} \boldsymbol{\beta_t} + \boldsymbol{\varepsilon_t}, & \text{if } t = 1,2m,3m, \ldots, T  \\
    \varnothing , & \text{otherwise}  \\
\end{cases}
\end{equation}
\begin{equation}
\begin{aligned}
\begin{bmatrix}
    \boldsymbol{y}_t \\
    \boldsymbol{\beta}_{t}  \\
    \end{bmatrix} =  \boldsymbol{\Phi}_0 + 
    \boldsymbol{\Phi}_1
    \begin{bmatrix}
    \boldsymbol{y}_{t-1} \\
    \boldsymbol{\beta}_{t-1}  \\
    \end{bmatrix} + \ldots +  
     \boldsymbol{\Phi}_p
    \begin{bmatrix}
    \boldsymbol{y}_{t-p} \\
    \boldsymbol{\beta}_{t-p}  \\
    \end{bmatrix} +
    \begin{bmatrix}
    \boldsymbol{u}_{y,t}  \\
    \tilde{u}_{q,t}   \\
    \end{bmatrix} \hspace{0.1cm},
\end{aligned} 
\end{equation}
where $m$ is the number of high-frequency observations contained within the span of one low-frequency observation period (for example in the case annual and quarterly observations $m = 4$). As both labor and capital are stock variables, the observation equation (\ref{obsstatespace}) assumes that the firm-level variables on labor and capital are reporting their end of the quarter values.\footnote{Accounting for flow variables in this framework would be more complicated, since the observed distribution would most likely be a convolution of the unobserved missing distributions as noted in \citet{marcellino2024nowcasting}.} To estimate the model we proceed again in two steps. First, we estimate the basis functions $\boldsymbol{H}$ using either principal component analysis or multilinear principal component analysis or CP decomposition. In particular, for a sufficient large number of observed low-frequency densities, we can still consistently estimate $\boldsymbol{H}$ by performing principal component analysis on the sample of observed low frequency densities. Then, we estimate the factor augmented VAR model using the Gibbs Sampler described in the previous section, which needs just to be adapted by considering the following observation equation
\begin{equation}\label{eq2mod}
    \boldsymbol{l} = \boldsymbol{M}\boldsymbol{b} + \boldsymbol{\epsilon} \hspace{0.1cm},  \hspace{1cm} \boldsymbol{\epsilon} \sim \mathcal{N}(\boldsymbol{0}, \sigma^2\boldsymbol{I}_{TN^{grid}}) \hspace{0.1cm},
\end{equation}
where now $\boldsymbol{M} = \boldsymbol{Q} \otimes \boldsymbol{H}$ where $\boldsymbol{Q} = \boldsymbol{I}_{T/m} \otimes diag(\boldsymbol{0}_{1 \times m - 1},1)$ is a $T \times T$ matrix selecting the factors according to the observation equation $(\ref{obsstatespace})$.  The posterior distribution of $\boldsymbol{b}$ is then just adjusted with this new definition of $\boldsymbol{M}$.

\section{Simulation from an heterogeneous firm model}\label{sec:simul}

In this section we assess the finite sample performance of the FunVAR model to recover the impulse response functions of the macroeconomic aggregates and the cross-sectional distributions to macroeconomic shocks  generated by a fully structural heterogeneous firms model. We consider the version of the standard heterogeneous firm model of \citet{KT2008} extended in \citet{winberry2018}.  Since the model by \citet{winberry2018}, once solved and approximated around the steady state, implies VAR dynamics in terms of the moments of the bivariate log-labor and log-capital distribution, our FunVAR model is inherently misspecified. Aware of this misspecification, we use this simulation study as evidence to assess whether and how well the FunVAR model can qualitatively replicate the effects of macroeconomic shocks on both the cross-sectional distributions and macroeconomic aggregates of the heterogeneous firm model. 

In Winberry's model, firms are subject to both idiosyncratic and aggregate productivity shocks. Firms are subject to heterogeneous adjustments costs. These costs play a critical role in the decision-making process across different firms regarding both production and input-allocation.\footnote{We refer to the paper of \citet{winberry2018} for further details on the model.} As a result of this decision-making process, the cross-sectional joint distribution of firm-level capital and labor evolves dynamically over time. Our specification of the FunVAR dynamics in terms of the joint distribution of labor and capital, rather than the marginal distributions, is designed to precisely capture the effects of structural shocks on the reallocation of both inputs. In other words, our model aims to capture the interdependence between these inputs in response to shocks, showing how firms adjust their allocation of labor and capital simultaneously. 

\begin{figure}[htp]
\caption{Approximation of $f(log(l),log(k))$ though bilinear principal components} \label{fig:sampleapprox}
\begin{center}
{\includegraphics[width=0.5\linewidth,height=0.3\textheight]{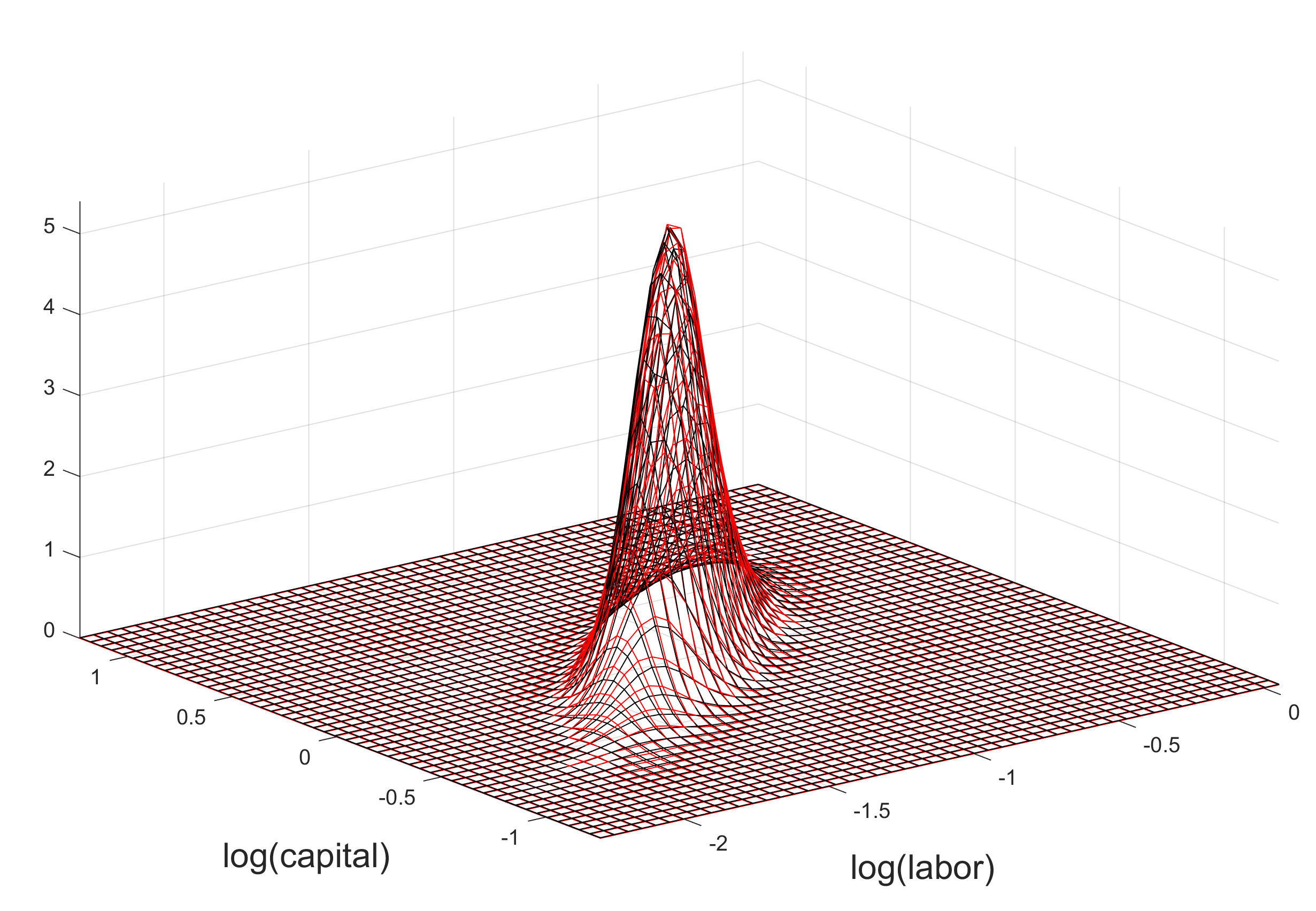}}  \\
\end{center}
\vspace*{-0.25cm}\parbox{1\textwidth}{\footnotesize{Notes: The figure shows the true (red) and the approximated (black) bivariate log-capital and log-labor  distribution from for one sample period in the simulation.}}
\end{figure}

We simulate the heterogeneous firm model for $T=250$ periods.  Table \ref{tab:calibration} in Appendix \ref{app_cali} reports details on the calibration of the parameters to simulate from the heterogeneous firm model, which directly follows \citet{winberry2018}. In general, we find both principal component analysis on the unfolded tensors and multilinear principal component analysis performing slightly better in terms of the approximation of the bivariate density. Figure \ref{fig:sampleapprox} shows one period approximation of the bivariate density using multilinear principal component analysis with $9$ basis functions, with $K_1=3$ and $K_2=3$. 

\begin{figure}[htp]
\caption{Responses to a TFP shock in the heterogeneous firm model} \label{fig:aggregresponse1}
\begin{center}
{\includegraphics[width=1\linewidth,height=0.4\textheight]{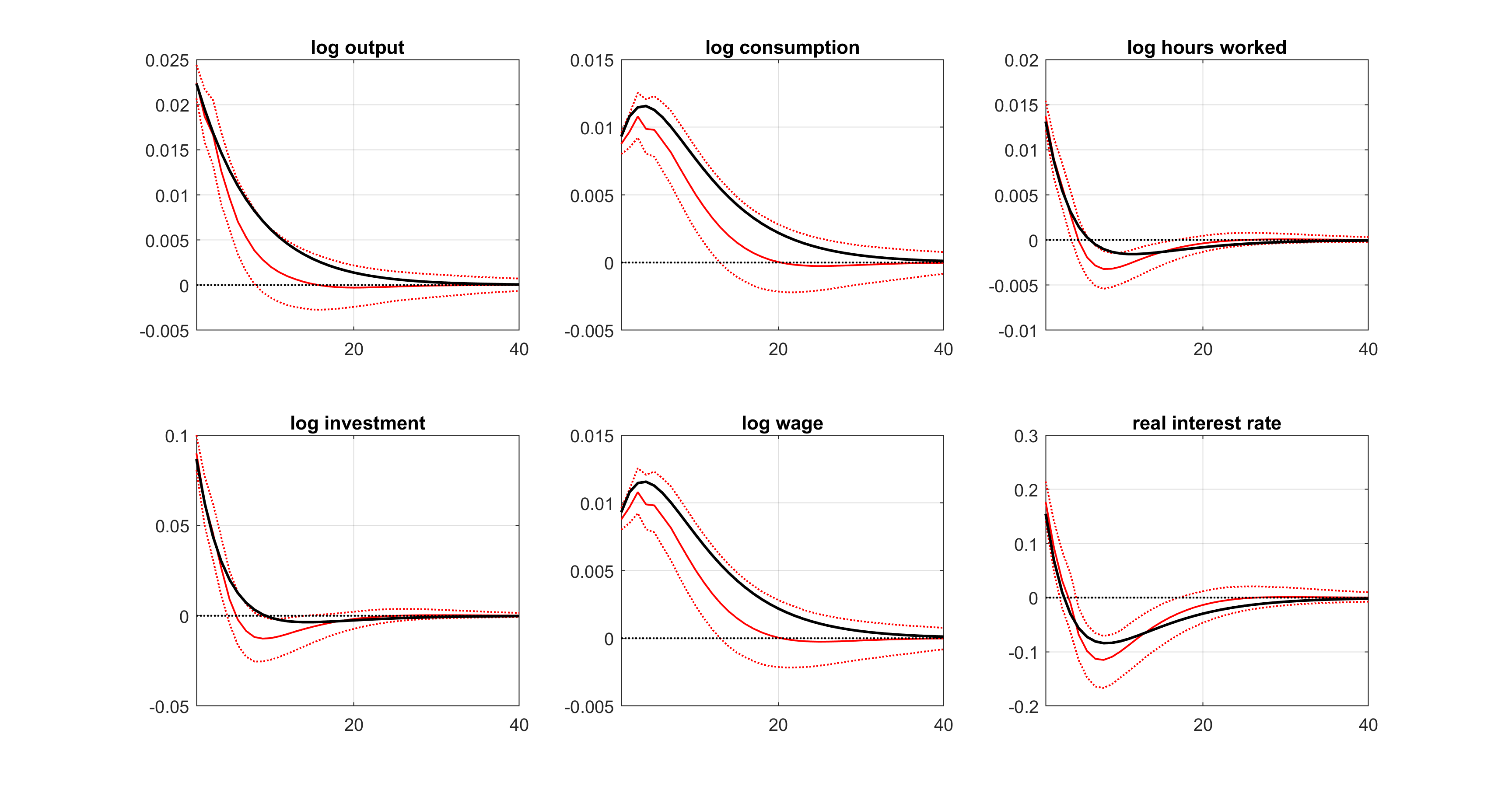}}  \\
\end{center}
\vspace*{-0.25cm}\parbox{1\textwidth}{\footnotesize{Notes: The figure shows the Impulse Response Functions (IRFs) of the macroeconomic aggregates to a one standard deviation TFP shock. In black we report the IRF of the \citet{winberry2018} heterogeneous firm model. In red bold line we report the posterior mean estimate from the FunVAR while in dashed red line we report the $5^{th}$ and $95^{th}$ credible sets. }}
\end{figure}

 We look at the estimated responses of the aggregate macroeconomic time series and the cross-sectional distributions to a one standard deviation aggregate TFP shock. Figure \ref{fig:aggregresponse1}, shows the impulse response functions of the main macroeconomic variables to a total factor productivity shock in the heterogeneous firm model. The figure reports in black the true responses. The red dashed lines are the $5^{th}$ and $95^{th}$ credible sets while
the solid red line is the posterior median estimate obtained from our FunVAR.  The TFP shocks are identified exploiting the exogeneity of the simulated TFP series. The model correctly recovers the dynamics of output, consumption, hours worked, investment and wages after the aggregate TFP shock hitting the economy. 

\begin{figure}[htp]
\caption{Responses of $f(\log(k), \log(l))$ to a TFP shock in the heterogeneous firm model} 
\label{fig:crossresponse}
\centering

\subfloat[$h = 4$ ]{
    \begin{minipage}{0.8\linewidth}
        \includegraphics[width=\linewidth, height=0.25\textheight]{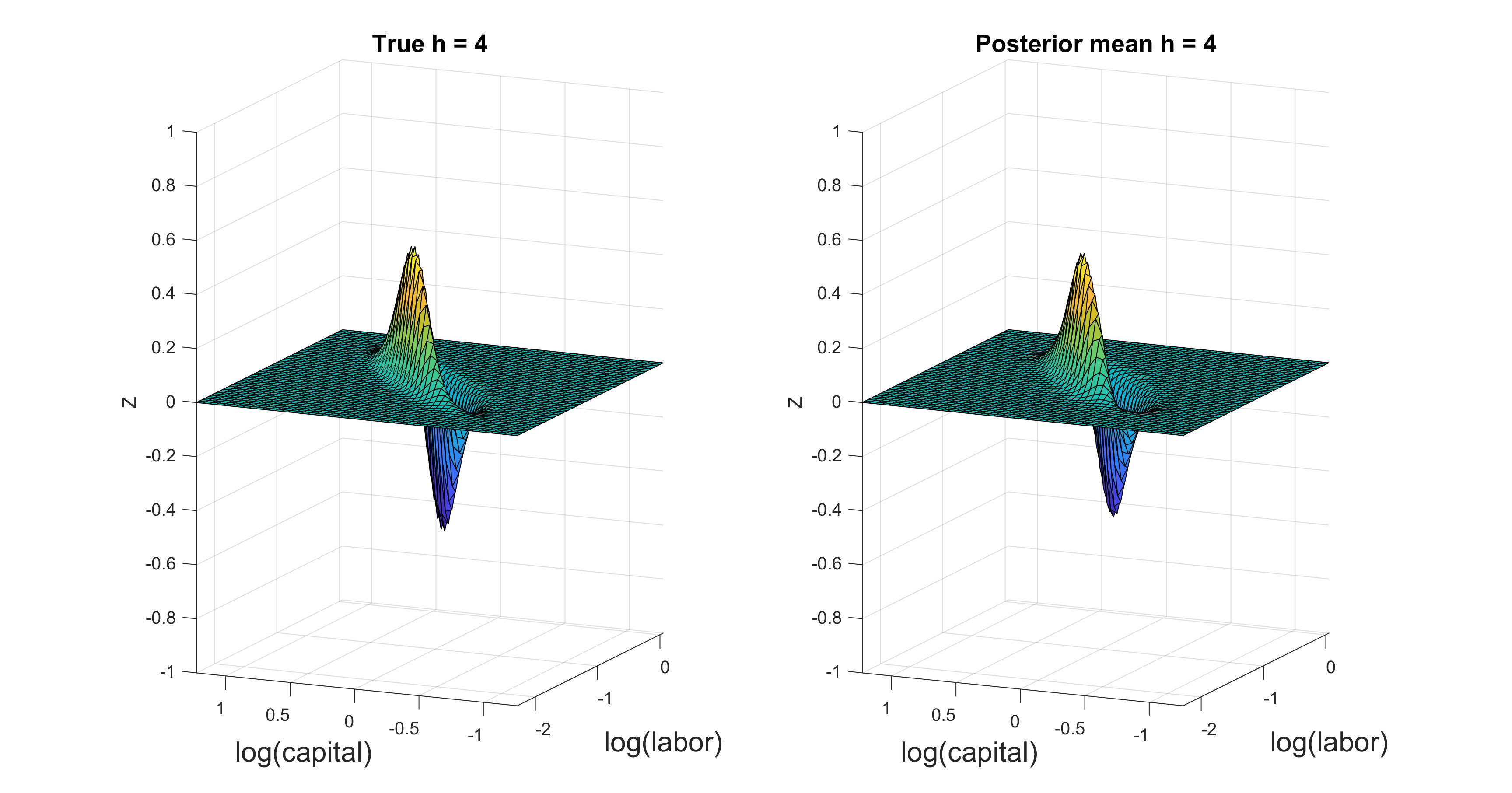}
    \end{minipage}
    \hfill
    \begin{minipage}{0.1\linewidth}
        \includegraphics[width=\linewidth]{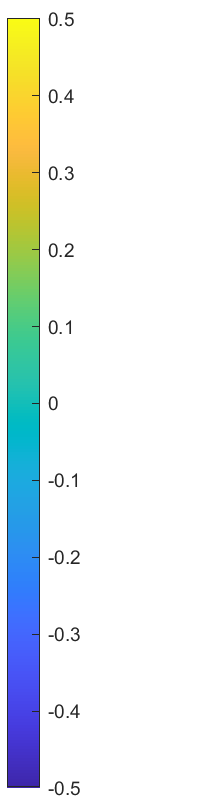}
    \end{minipage}
} \\
\subfloat[$h = 8$ ]{
    \begin{minipage}{0.8\linewidth}
        \includegraphics[width=\linewidth, height=0.25\textheight]{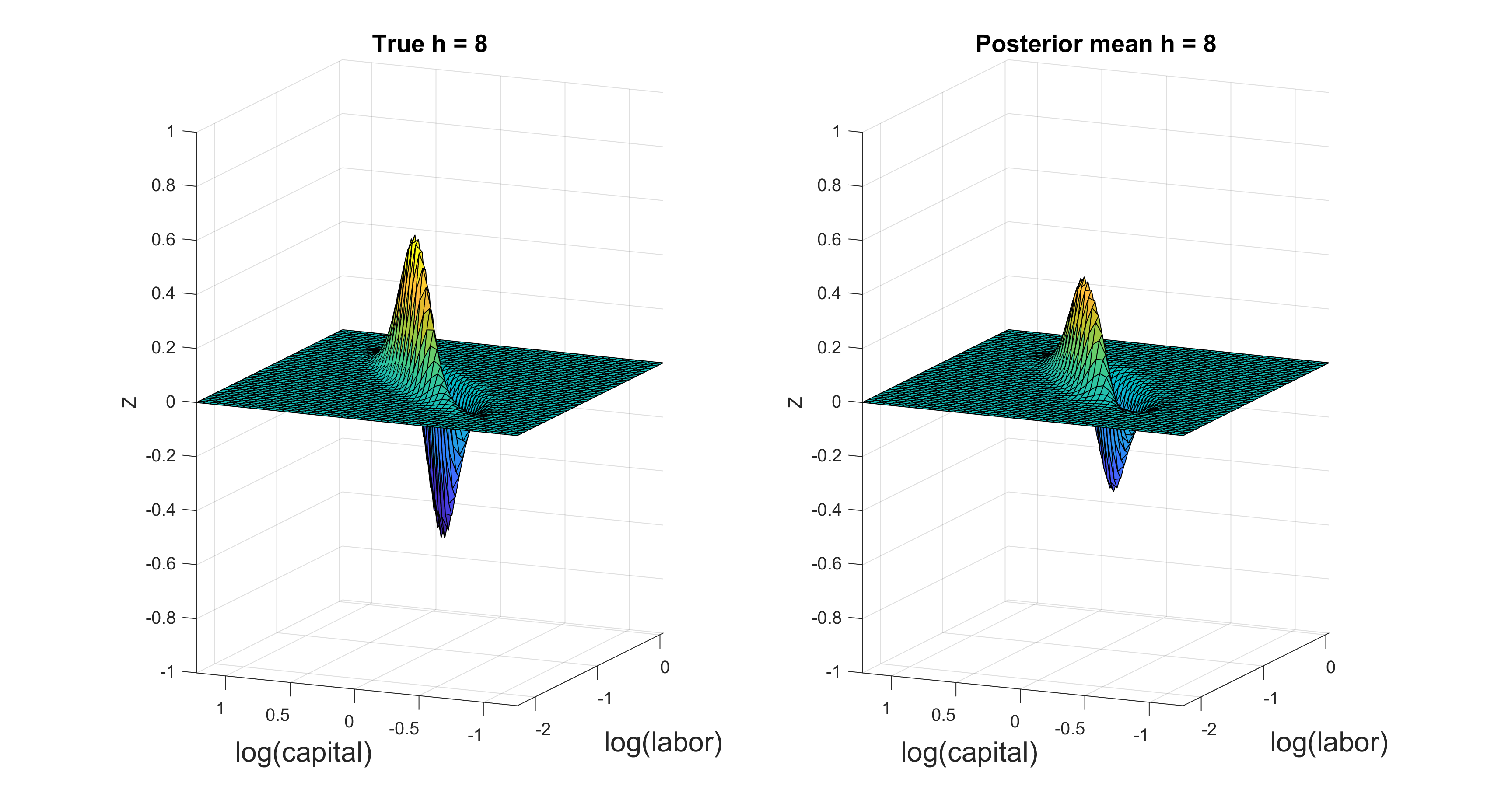}
    \end{minipage}
    \hfill
    \begin{minipage}{0.1\linewidth}
        \includegraphics[width=\linewidth]{colorbar_only.png}
    \end{minipage}
} \\
\subfloat[$h = 24$ ]{
    \begin{minipage}{0.8\linewidth}
        \includegraphics[width=\linewidth, height=0.25\textheight]{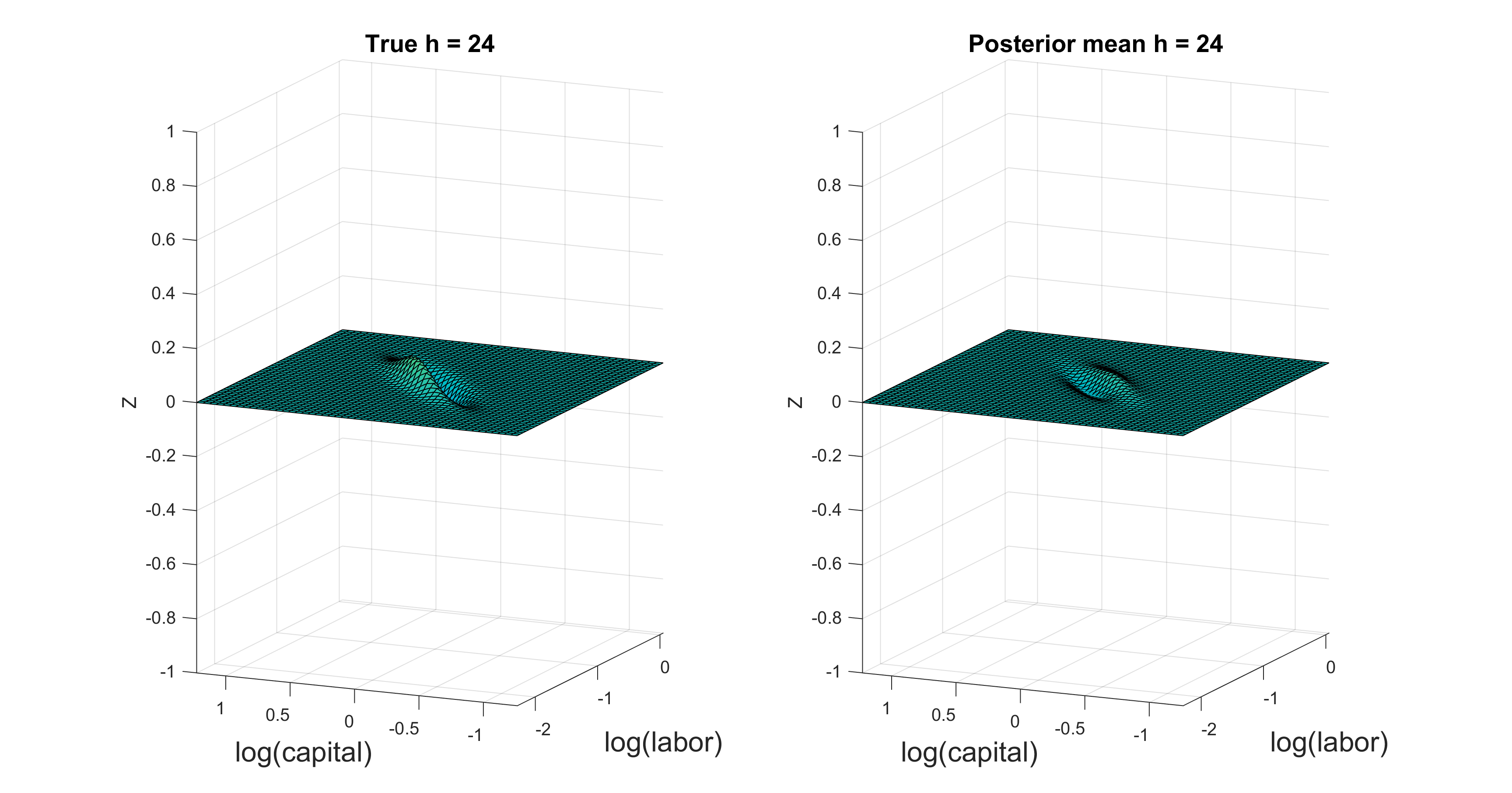}
    \end{minipage}
    \hfill
    \begin{minipage}{0.1\linewidth}
        \includegraphics[width=\linewidth]{colorbar_only.png}
    \end{minipage}
}

\vspace*{-0.2cm}
\parbox{1\textwidth}{\footnotesize{Notes: The figure shows the bivariate FIRFS for 4,8 and 24 periods following the TFP shock. \\ On the left hand side the true FIRFS, while on the right hand side the posterior mean estimate from the FunVAR.}}
\end{figure}

Figure \ref{fig:crossresponse} reports the Functional Impulse Response Functions (FIRF) of the bivariate distribution of firm-level capital and labor. FIRFs are obtained by computing the difference in the mass between the bivariate log-capital and log-labor distributions after the TFP shock has occurred and the steady state distribution (this is reported in the z axis). We show the FIRF after 4 periods (one year), 8 periods (2 years) and 24 periods (6 years). In the figures in the left panels, we show the true FIRF, while in those on the right panels we report the posterior median FIRF estimated with the FunVAR on the simulated data. The posterior mean estimate from the FunVAR model correctly tracks the evolution of the bivariate density after the aggregate TFP shock hitting the economy. 

\begin{figure}[htp]
    \centering
    \includegraphics[width=0.8\linewidth]{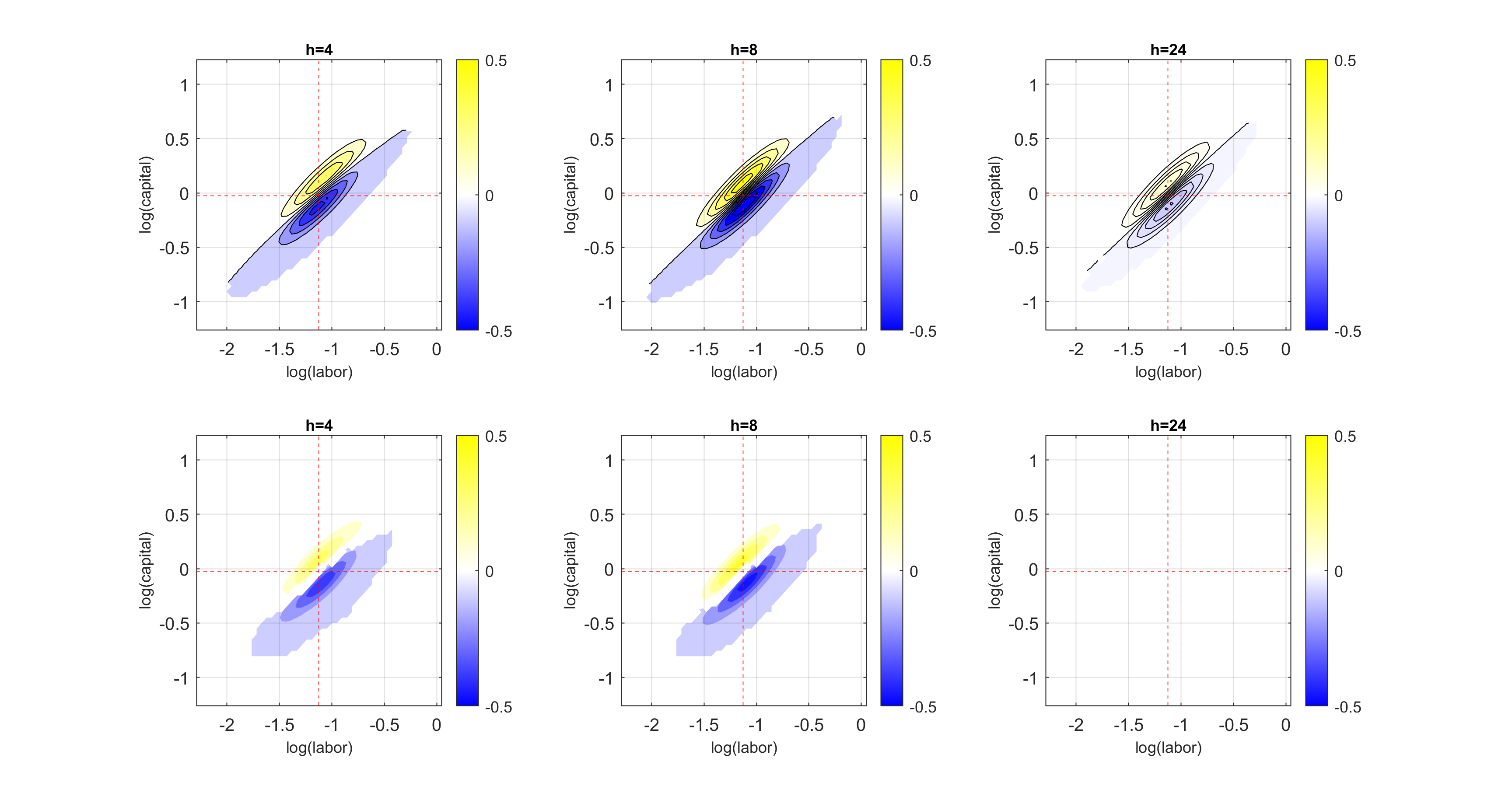}
    \caption{Contours from the bivariate FIRF of $(\log(l),\log(k))$}   
    \label{fig:contours_simulation}
    
    \vspace*{-0.2cm}
    \parbox{1\textwidth}{\footnotesize{Notes: The figure shows the contours from the bivariate FIRFs for 4,8 and 24 periods following the TFP shock. The panel above reports the true FIRF from the heterogeneous firm model, while the panel below reports value for the posterior mean estimate from the FunVAR only if the $15^{\text{th}}-85^{\text{th}}$ credible set does not contain zero. Dashed red lines are the steady state mean values in the heterogeneous firm model.}}
\end{figure}

As we are also concerned about whether the effects of a TFP are efficiently estimated, in Figure \ref{fig:contours_simulation} we report the contours of the bivariate FIRF. The upper panels report the contours from the true FIRF in the heterogeneous firm model. The lower panels instead  report the posterior median value, only in the case the 65 \% credible regions of the posterior distribution do not contain zero. The figure shows that the effects of the TFP shocks are precisely estimated, and in general more accurately estimated at lower horizons. Overall, this exercise with simulated data shows that the FunVAR is quite reliable also in finite samples to recover the distributional effects of TFP shocks.

\section{The distributional effects of  effects of TFP shocks in the U.S.}\label{sec:emp_appl}
We now exploit the FunVAR model to examine the effects of TFP shocks on the US economy. An extensive literature dating back to \citet{kydland1982time} has explored the role of TFP shocks in driving business cycle fluctuations.  This ongoing debate underscores the complexity of understanding how technology shocks affect firms allocation of capital and labor inputs. We use the FunVAR to investigate the effects of TFP shocks both on the macroeconomic aggregates and the firm-level joint capital and labor distribution. 

\subsection{Aggregate and firm-level data}
Our analysis leverages the Compustat dataset to extract the cross-sectional distribution of U.S. firm-level labor and capital. The Annual Compustat dataset provides joint data on firm-level capital and employees, while firm-level capital data is also available at a quarterly frequency through the Quarterly Compustat dataset. In what follows, we present results based a quarterly dataset obtained by micro-level interpolation  to estimate missing intra-year labor observations. An alternative approach is to rely on the annual micro-level dataset for both capital and labor, and use the state-space framework outlined in Section \ref{mixedfrequency} to combine the yearly distributions with the quarterly aggregate macroeconomic time series. More specifically, we consider cross sectional data on firm's balance sheets from 1984-Q4 to 2019-Q4. The raw data are cleaned and transformed mostly following \citet{ottonoellowinberry}, as detailed in Appendix \ref{data_transformation}. The final dataset is comprised of cross sections with an average size of 2.809 firms. Since balance sheet data are expressed in nominal terms (millions of dollars), we obtain the value of real capital dividing by the implicit price deflator of the nonresidential gross private domestic fixed investment. In order to clean the micro data from low frequency fluctuations and concentrate on cyclical fluctuations,  we log-linearly-detrend the level of capital and labor at the firm level. Concerning the data on the macroeconomic aggregates, these are taken from FRED-QD (Federal Reserve Economic Data Quarterly Dataset). We consider real gross domestic product (\texttt{GDPC1}), real personal consumption expenditures (\texttt{PCECC96}), average yearly hours worked (\texttt{AWHNONAG}), nonresidential real private fixed investment (\texttt{PNFIx}) and real hourly non-farm business sector compensation (\texttt{COMPRNFB}) and the real interest rate (obtained as \texttt{FEDFUND} - $\pi^{\texttt{CPIAUSL}}$).\footnote{$\pi^{\texttt{CPIAUSL}}$ is defined as the yearly inflation rate obtained from the \texttt{CPIAUSL} price series.} All the variables, excluding the real interest rate,  are log-linearly detrended. 
\subsection{Aggregate and disaggregated response to TFP shocks}
\begin{figure}[htp]
\caption{Responses to a TFP shock in the heterogeneous firm model} \label{fig:aggregresponse}
\begin{center}
{\includegraphics[width=1\linewidth,height=0.4\textheight]{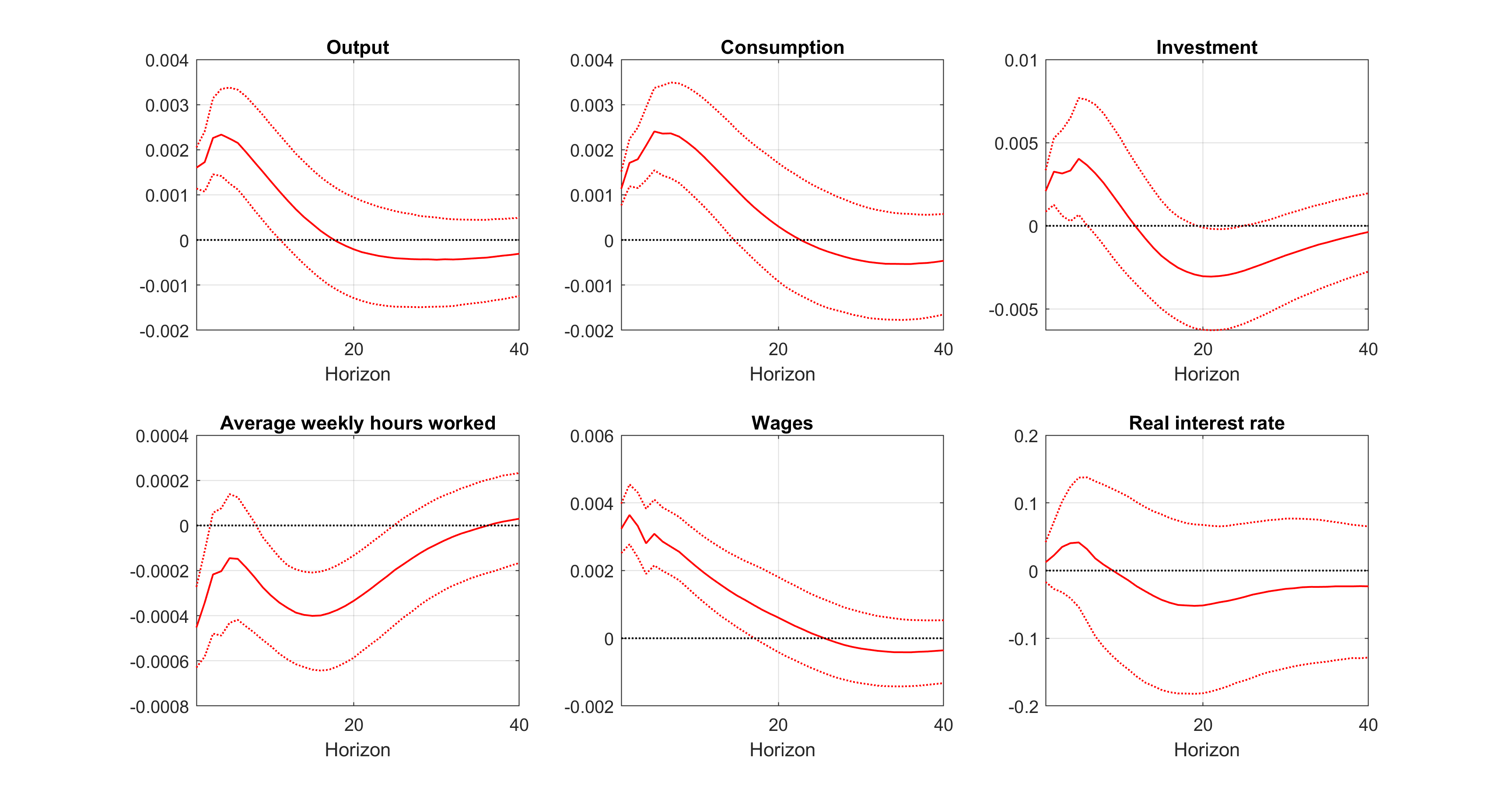}}  \\
\end{center}
\vspace*{-0.25cm}\parbox{1\textwidth}{\footnotesize{Notes: The figure shows the Impulse Response Functions (IRFs) of the macroeconomic aggregates to a one standard deviation TFP shock. In red bold line we report the posterior mean estimate from the FunVAR while in dashed red line we report the $15^{th}$ and $85^{th}$ credible sets. }}
\end{figure}

We estimate the functional VAR model on the sample of US data that goes from 1984-Q4 to 2019-Q4. The VAR comprises seven macroeconomic variables, being the \citet{fernald2014quarterly} TFP series, real personal consumption expenditures, average yearly hours worked, nonresidential real private fixed investment, real hourly non-farm business sector compensation and real interest rate. In what follows, we present evidences based on the identification of the aggregate TFP shock trough the internal instrument procedure, exploiting the TFP measure developed in \citet{fernald2014quarterly} as our exogenous proxy for the TFP shocks. More in detail, we assume that the TFP proxy is contemporaneously exogenous with respect to the other variables entering the VAR. We exploit the Cholesky decomposition of the variance covariance matrix $\boldsymbol{\Sigma}$ to identify the column of the impact matrix of the structural VAR, which relates to the effect of the TFP shock both on the aggregate macroeconomic variables and the bivariate labor and capital distributions. Figure \ref{fig:aggregresponse} shows the response of the aggregate macroeconomic variables to the aggregate TFP shock. The response of the macroeconomic aggregates aligns with the responses in the real business cycle model, with the exception of hours worked. In fact, while in the heterogeneous firm real business cycle model a positive TFP shock leads to an increase on impact of hours worked, we find that a positive TFP shock leads to a persistent decrease of hours worked, as previously found in other studies \citep{gali1999technology,fernald2014quarterly,basu2006are}.

\begin{figure}[H]
    \centering
    \begin{subfigure}{\textwidth}
        \centering
        \includegraphics[width=0.95\textwidth]{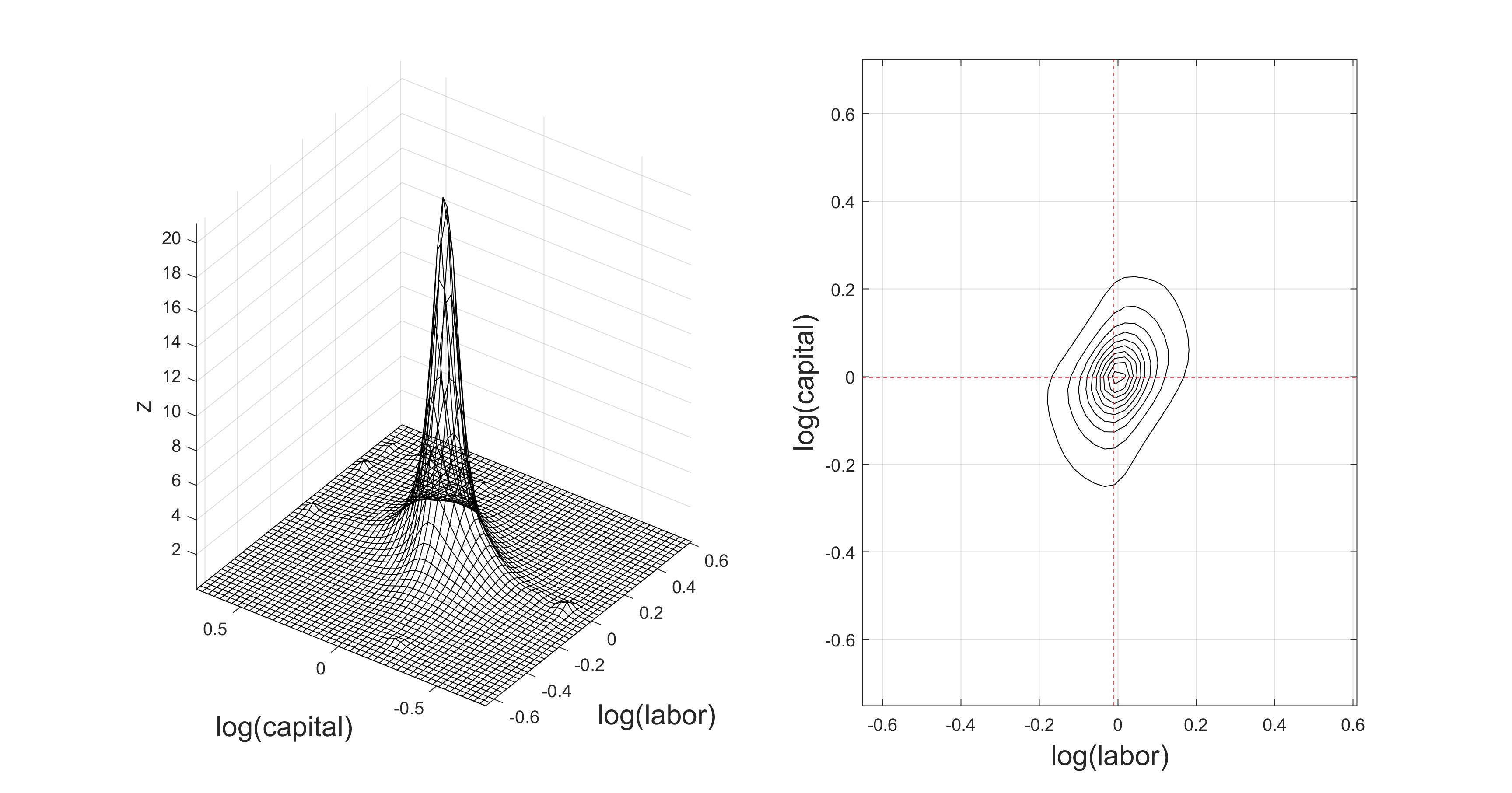}
        \caption{Posterior mean steady state distribution of $\log(l),\log(k)$.}
        \label{fig:left_image1}
    \end{subfigure}
    \hfill
    \begin{subfigure}{\textwidth}
        \centering
        \begin{subfigure}{0.48\textwidth}
            \centering
            \includegraphics[width=\textwidth]{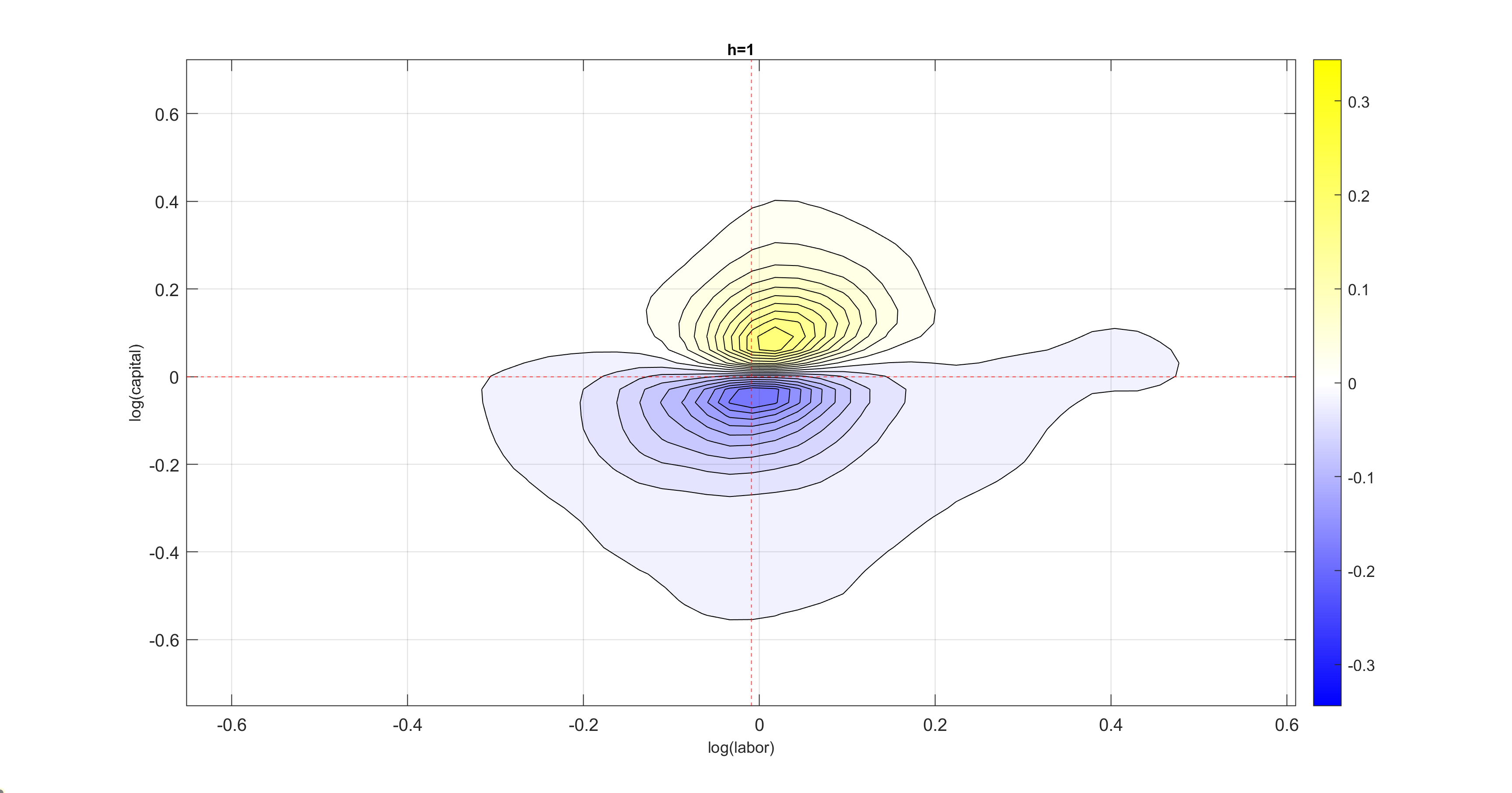}
            \caption*{$h=1$}
        \end{subfigure}
        \hfill
        \begin{subfigure}{0.48\textwidth}
            \centering
            \includegraphics[width=\textwidth]{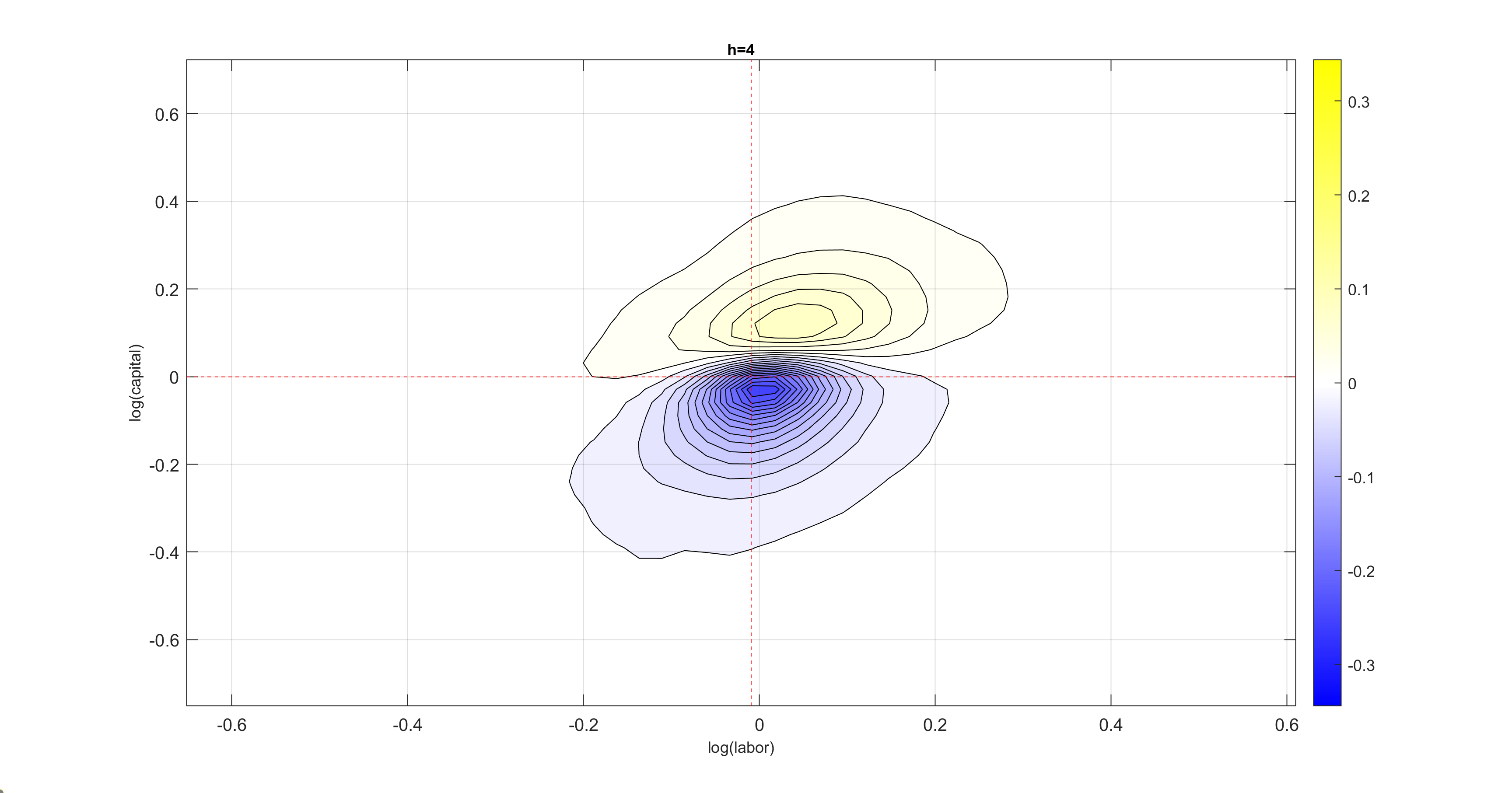}
            \caption*{$h=4$}
        \end{subfigure}
        \vskip\baselineskip 
        \begin{subfigure}{0.48\textwidth}
            \centering
            \includegraphics[width=\textwidth]{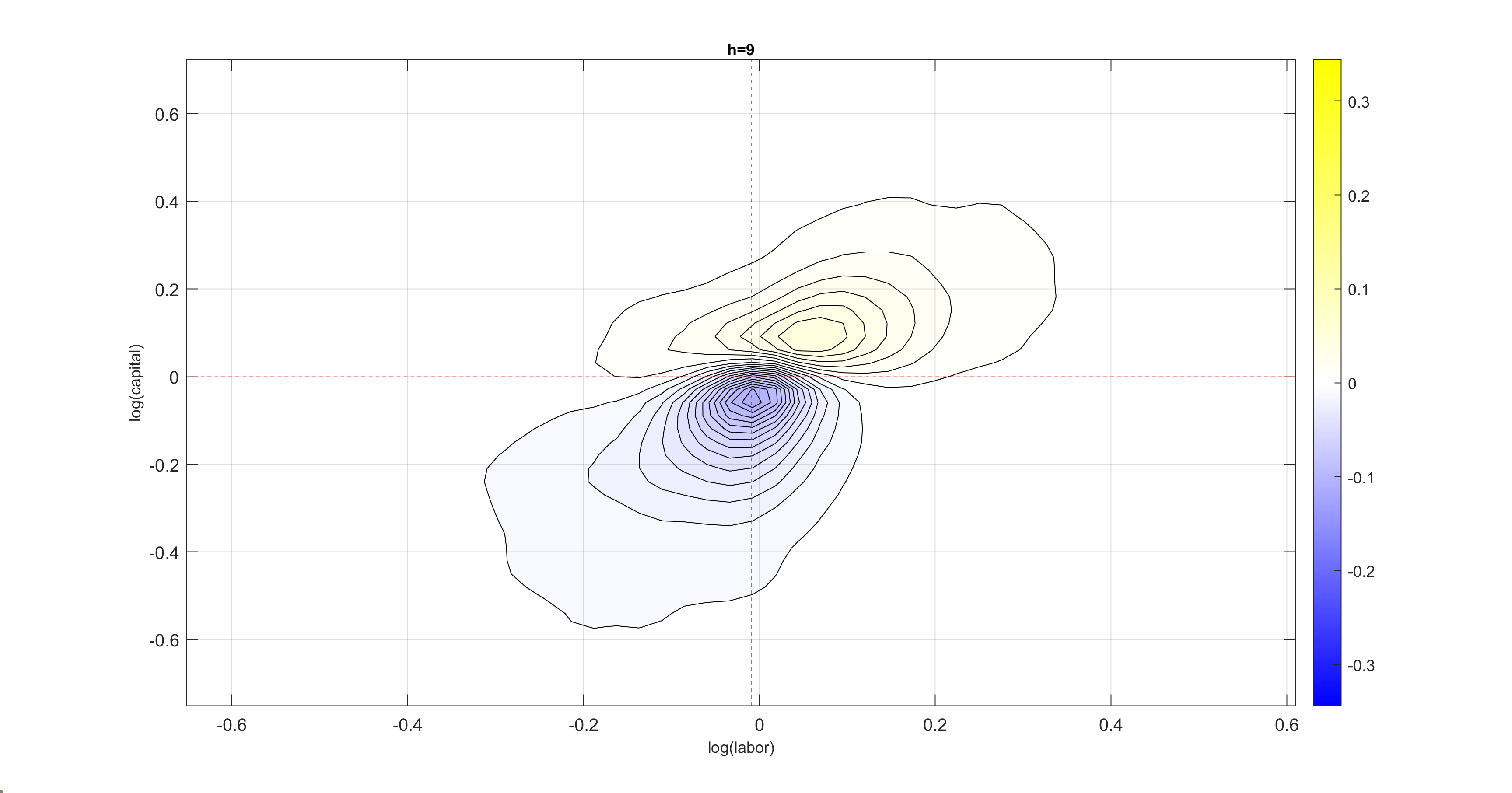}
            \caption*{$h=8$}
        \end{subfigure}
        \hfill
        \begin{subfigure}{0.48\textwidth}
            \centering
            \includegraphics[width=\textwidth]{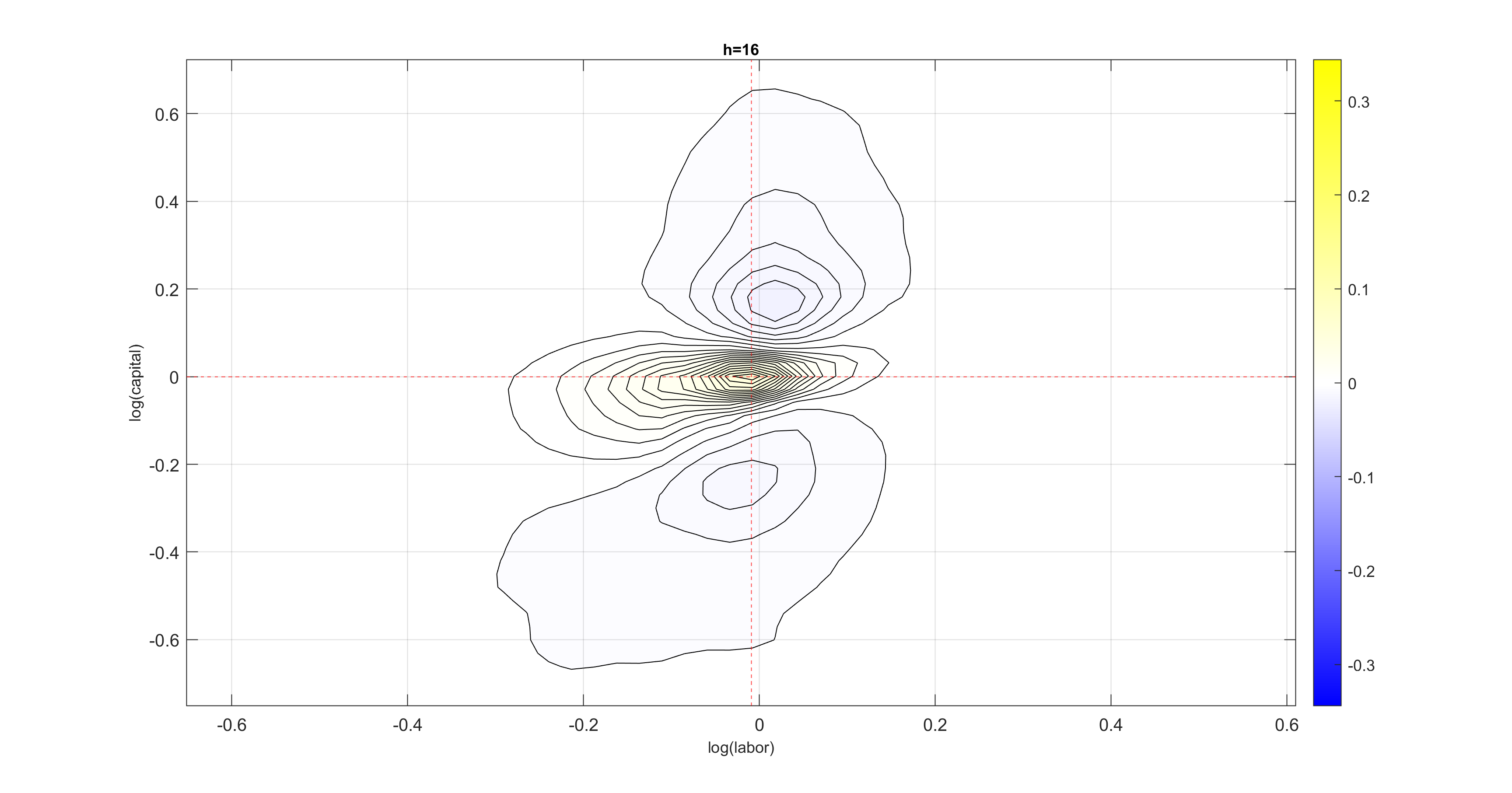}
            \caption*{$h=16$}
        \end{subfigure}
        \caption{Contours of the posterior mean estimate of the bivariate FIRF following an aggregate TFP shock.}
        \label{fig:right_image}
    \end{subfigure}
    
    \caption{Response of the cross sectional distributions to an aggregate TFP shock}\label{fig:combined_images2}
    
    \vspace*{0.15cm}
    \parbox{1\textwidth}{\footnotesize{Notes: The panel above shows the estimated posterior mean steady state distribution of $\log(l)$ and $\log(k)$. The panel below shows the contours of the posterior mean estimate of the bivariate FIRF following an aggregate TFP shock for 1, 4, 8, and 16 periods ahead. The dashed line reports the posterior mean steady state values.}}
\end{figure}

We use the FunVAR to study the micro-level propagation of the TFP shock. In particular, we look at the effect of the TFP shock on the bivariate distribution of labor and capital across firms. Figure \ref{fig:combined_images2}
reports the posterior median estimate of the steady state distribution of log-labor and log-capital (left upper panel) with the corresponding  contour levels (right upper panel). The lower panels report the contour plots relative to the change in the mass w.r.t the steady state value in the bivariate density of labor and capital, for different horizons after the TFP shock has occurred. It turns out that some firms expand both capital and labor simultaneously, indicating that they are growing and scaling their operations in response to the productivity gains from the shock. This is reflected by the increase of the mass of firms with both capital and labor above theirs steady state level. However, the figure also shows that after the TFP shock, the mass of firms with capital above the steady state level  (red dashed lines) increases, for all levels of labor. While the mass of firms with capital endowments greater than the steady state level and labor below the steady state level increases, the mass of firms with labor above the steady state and capital below the steady state level decreases unambiguously. 
After 16 quarters (four years), the effects of the shock gets reabsorbed. 


\section{Conclusion}\label{sec:concl}
We develop a Functional Augmented Vector Autoregression (FunVAR) model to explicitly incorporate firm-level heterogeneity observed in more than one dimension and study its interaction with aggregate macroeconomic fluctuations. We remark the importance of modeling the joint distribution, rather than just the marginal distributions of the micro observables, for qualitatively assessing the effects of economic shocks on the micro-level distributions and for determining which scenario, potentially aligned with a specific structural model, the data support more. We address the challenge of approximating a multidimensional distribution using data reduction techniques for tensor data objects. We use these methods for approximating multiple cross-sectional distributions of micro-variables, which dynamically interact with the macroeconomic aggregates in a Functional VAR framework. We use the model to study the transmission of aggregate TFP shocks both on the cross-sectional distribution of firm-level capital and labor and the macroeconomic aggregates.  We find that the shock causes a shift where fewer firms combine higher labor with lower capital, while more firms accumulate both capital and labor above steady-state levels.

\printbibliography
\clearpage
\appendix
\section{Appendix}
\subsection{Factor augmented approximation of the Functional VAR model}\label{sec_approx}
Thanks to the finite dimensional approximation (\ref{finite_approx}), and taking $\boldsymbol{s}_{K}(\boldsymbol{x'})$ as a $K$-dimensional vector of functional basis such that  
\begin{equation}
    \int  \boldsymbol{s}_{K}(\boldsymbol{x}) \boldsymbol{h}_K(\boldsymbol{x})'  d \boldsymbol{x}= \boldsymbol{C_{\beta}}\hspace{0.1cm},
\end{equation}
we can rewrite the function augmented VAR expanding the unknown functions of $\boldsymbol{x}$ as follows:
\begin{equation}
\begin{aligned}
    c_{l}(\boldsymbol{x}) &= \boldsymbol{h}_K(\boldsymbol{x})'\tilde{c}_{l,t} \\
    \boldsymbol{B}_{s,yl}(\boldsymbol{x}) &= \boldsymbol{B_{s,yl}}\boldsymbol{s}_{K}(\boldsymbol{x}) \\ 
    \boldsymbol{B}_{s,ly}(\boldsymbol{x}) &= \boldsymbol{h}_K(\boldsymbol{x})'\boldsymbol{B_{s,ly}} \\  
    B_{s,ll}(\boldsymbol{x},\boldsymbol{x'}) &= \boldsymbol{h}_K(\boldsymbol{x})'\boldsymbol{B_{s,ll}}\boldsymbol{s}_{K}(\boldsymbol{x'}) \\
    \boldsymbol{u}_{l}(\boldsymbol{x}) &= \boldsymbol{h}_K(\boldsymbol{x})'\tilde{u}_{l,t} \hspace{0.1cm},
\end{aligned}
\end{equation}
where $\boldsymbol{B_{s,ll}}$ is a $K \times K$ matrix, $\boldsymbol{B_{s,ly}}$ is a $K \times n_y$ matrix, and $\boldsymbol{B_{s,yl}}$ is an $n_y \times K$ matrix. Plugging in (\ref{fvar}) and (\ref{fvar2}) we get:

\begin{equation}\label{fvar11}
    \boldsymbol{y_t} = \boldsymbol{c_y} + \sum_{s=1}^p \boldsymbol{B}_{s,yy}\boldsymbol{y_{t-s}} + \sum_{s=1}^p  \boldsymbol{B_{s,yl}}\boldsymbol{C_{\beta}} \boldsymbol{\beta}_ {t-s;K} + \boldsymbol{u_{y,t}}
\end{equation}
\begin{equation}
    \boldsymbol{h}_K(\boldsymbol{x})'\boldsymbol{\beta}_{t} =  \boldsymbol{h}_K(\boldsymbol{x})'\tilde{c}_{l,t} + \sum_{s=1}^p \boldsymbol{h}_K(\boldsymbol{x})'\boldsymbol{B_{s,ly}}\boldsymbol{y}_{t-s} + \boldsymbol{h}_K(\boldsymbol{x})'\sum_{s=1}^P \boldsymbol{B_{s,ll}}\boldsymbol{\beta}_{t-s;K} + \boldsymbol{h}_K(\boldsymbol{x})'\tilde{u}_{l,t} \hspace{0.1cm},       
\end{equation}
which becomes:

\begin{equation}\label{fvar22}
    \boldsymbol{\beta}_{t} =  \tilde{c}_{l,t} + \sum_{s=1}^p \boldsymbol{B_{s,ly}}\boldsymbol{y}_{t-s} +  \sum_{s=1}^P \boldsymbol{B_{s,ll}}\boldsymbol{C_{\beta}}\boldsymbol{\beta}_{t-s;K} + \tilde{u}_{l,t} \hspace{0.1cm}.     
\end{equation}

\subsection{Details on Compustat data}\label{data_transformation}
We mostly follow \citet{ottonoellowinberry} for what concerns data cleaning and variable transformations. 
In our analysis, we initialize the assessment of each firm's investment behavior by setting the initial value of capital stock for the next period, \( k_{j,t+1} \), based on the reported level of gross plant, property, and equipment from Compustat (\texttt{ppegtq}, item 118). We continue by tracking the evolution of this capital stock through changes in net plant, property, and equipment (\texttt{ppentq}, item 42), which usually offers more observations and incorporates adjustments for depreciation. When encountering missing data points in \texttt{ppentq} between two periods, linear interpolation is employed using the nearest available values to estimate the missing entry. However, we avoid interpolation when the dataset exhibits two or more consecutive missing observations to maintain accuracy. Our empirical analysis implements several exclusion criteria to ensure data quality and relevance. Initially, we exclude firms operating within specific sectors deemed non-representative of typical corporate investment behaviors. These sectors include finance, insurance, and real estate, which are categorized within Standard Industrial Classification (SIC) codes ranging from 6000 to 6799, as well as utilities with SIC codes from 4900 to 4999. Additionally, non-operating establishments (SIC code 9995) and industrial conglomerates (SIC code 9997) are also omitted from our study. Moreover, our analysis is restricted to firms that are incorporated in the United States, excluding any firm-quarter observations that fail to meet this criterion. Within the retained data, we further refine our sample by excluding observations that display characteristics of extreme financial behavior or data anomalies. This includes observations with negative capital or assets, and those involving significant acquisitions, defined as acquisitions where the acquired assets are greater than 5\% of the firm's total assets. We also exclude observations where the investment rate appears anomalously high or low, falling in the top or bottom 0.5\% of the distribution, as well as those where the investment spell is shorter than 40 quarters. Additional exclusions apply to quarters with real sales growth exceedingly high above 1 or significantly low below -1. Observations with negative sales or liquidity are similarly omitted to maintain the integrity and reliability of our analysis. 
We transform the nominal firm level capital into real firm level capital dividing by the implicit price deflator of the nonresidential gross private domestic fixed investment available from FRED (\texttt{A008RD3Q086SBEA}). We log-linearly detrend the level of capital at the firm level to concentrate on cyclical fluctuations. For the labor variable we consider \texttt{emp} from the Annual dataset and interpolate quarterly observations from annual observations.  
\clearpage
\subsection{Calibration of the heterogeneous firm model in the simulation}\label{app_cali}
\begin{table}[h]
\centering
\begin{tabular}{@{}ll@{}}
\toprule
\toprule
\textbf{Parameter} & \textbf{Value} \\ \midrule
$\beta$ (Discount factor) & 0.961 \\
$\sigma$ (Utility curvature) & 1 \\
$\alpha$ (Inverse Frisch limit $\alpha \to 0$) &  \\
$\xi$ (Fixed cost) & 0.0083 \\
$\chi$ (Labor disutility) & N/A (ensures $N^* = 1/3$) \\
$\nu$ (Labor share) & 0.64 \\
$\theta$ (Capital share) & 0.256 \\
$\delta$ (Capital depreciation) & 0.085 \\
$\rho_z$ (Aggregate TFP AR(1)) & 0.859 \\
$\sigma_z$ (Aggregate TFP AR(1) std. dev.) & 0.014 \\
$\rho_\epsilon$ (Idiosyncratic TFP AR(1)) & 0.859 \\
$\sigma_\epsilon$ (Idiosyncratic TFP AR(1) std. dev.) & 0.022 \\
$a$ (No fixed cost region) & 0.011 \\
\bottomrule
\bottomrule
\end{tabular}
\caption{ Parameterization of the heterogeneous firms model}\label{tab:calibration}
\end{table}

\end{document}